\documentclass[12pt]{iopart}

\usepackage{iopams} \usepackage{graphics} \usepackage{bm}

\def\eq{&=&} \def\nn{\nonumber}

\begin{document}

\topical[]{Wigner crystal physics in quantum wires}

\author{Julia S Meyer} \address{Department of Physics, The Ohio State
  University, Columbus, Ohio 43210, USA} \author{K A Matveev}
\address{Materials Science Division, Argonne National Laboratory,
  Argonne, Illinois 60439, USA}

\ead{jmeyer@mps.ohio-state.edu}
\begin{abstract}
  The physics of interacting quantum wires has attracted a lot of
  attention recently. When the density of electrons in the wire is
  very low, the strong repulsion between electrons leads to the
  formation of a Wigner crystal. We review the rich spin and orbital
  properties of the Wigner crystal, both in the one-dimensional and
  quasi-one-dimensional regime. In the one-dimensional Wigner crystal
  the electron spins form an antiferromagnetic Heisenberg chain with
  exponentially small exchange coupling. In the presence of leads the
  resulting inhomogeneity of the electron density causes a violation
  of spin-charge separation. As a consequence the spin degrees of
  freedom affect the conductance of the wire. Upon increasing the
  electron density, the Wigner crystal starts deviating from the
  strictly one-dimensional geometry, forming a zigzag structure
  instead. Spin interactions in this regime are dominated by ring
  exchanges, and the phase diagram of the resulting zigzag spin chain
  has a number of unpolarized phases as well as regions of complete
  and partial spin polarization. Finally we address the orbital
  properties in the vicinity of the transition from a one-dimensional
  to a quasi-one-dimensional state.  Due to the locking between chains
  in the zigzag Wigner crystal, only one gapless mode exists.
  Manifestations of Wigner crystal physics at weak interactions are
  explored by studying the fate of the additional gapped low-energy
  mode as a function of interaction strength.
\end{abstract}


\submitto{\JPCM}

\maketitle

\section{Introduction}

First experiments~\cite{Wees,Wharam} on electronic transport in
one-dimensional conductors revealed the remarkable quantization of
conductance in multiples of the universal quantum $2e^2/h$, where $e$
is the elementary charge and $h$ is Planck's constant.  These
experiments were performed by confining two-dimensional electrons in
GaAs heterostructures to one dimension by applying a negative voltage
to two gates, thereby forcing the electrons to flow from one side of
the sample to the other via a very narrow channel.  Such devices,
typically referred to as quantum point contacts, are the simplest
physical realization of a one-dimensional electron system.  Although
the length of the one-dimensional region in quantum point contacts is
relatively short, the quantization of conductance indicates that
transport in such devices is essentially one-dimensional.  Longer
quantum wires have been created later using either a different gate
geometry \cite{Tarucha}, or by confining two-dimensional electrons by
other means, such as in cleaved-edge-overgrowth devices \cite{Yacoby}.
Finally, a fundamentally different way of confining electrons to one
dimension has been recently realized in carbon nanotubes
\cite{Tans,Bockrath}.  The interest in the study of one-dimensional
conductors is stimulated by the relatively low disorder in these
systems and by the ability to control their parameters.  For instance,
the effective strength of the electron-electron interactions is
determined by the electron density, which can be tuned by changing the
gate voltage.  Thus quantum wire devices represent one of the simplest
interacting electron systems in which a detailed study of transport
properties can be performed.

Interactions between one-dimensional electrons are of fundamental
importance.  Unlike in higher-dimensional systems, in one dimension
the low-energy properties of interacting electron systems are not
described by Fermi-liquid theory.  Instead, the so-called
Tomonaga-Luttinger liquid emerges as the proper description of the
system in which, instead of fermionic quasiparticles, the elementary
excitations are bosons \cite{giamarchi}.  Interestingly, the
quantization of conductance in quantum point contacts is well
understood in the framework of noninteracting electrons
\cite{Landauer} despite the relatively strong interactions in these
devices.  This paradox was resolved theoretically \cite{maslov,
  ponomarenko, safi} by considering a Luttinger liquid with
position-dependent parameters chosen in a way that models strongly
interacting electrons in the quantum wire connected to leads in which
interactions can be neglected.  It was found that the dc conductance
of such a system is completely controlled by the leads, and is
therefore insensitive to the interactions.

The latter conclusion is in apparent disagreement with experiments
observing the so-called \emph{0.7 structure} in the conductance of
quantum point contacts \cite{Thomas, Thomas_1, Thomas_2, Crook,
  Kristensen, Kane, Reilly_1, Cronenwett, Rokhinson}.  This feature
appears as a quasi-plateau of conductance at about $0.7\times 2e^2/h$
at very low electron density in the wire, and usually grows with
temperature.  A number of possible explanations have been proposed,
most of which attribute the feature to the fact that at low densities
the effective interaction strength is strongly enhanced.  One of the
most common explanations attributes the 0.7 structure to spontaneous
polarization of electron spins in the wire \cite{Thomas, Thomas_1,
  Thomas_2, Crook, Kristensen, Kane, Reilly_1, Rokhinson, Berggren1,
  Berggren2, Berggren3, Spivak, Reilly}.  Although such polarization
is forbidden in one dimension \cite{Lieb}, the electrons in quantum
wires are, of course, three-dimensional, albeit confined to a channel
of small width.  This deviation from true one-dimensionality may, in
principle, give rise to a spin-polarized ground state of the
interacting electron system.

\begin{figure}[tb]
  \hfill \resizebox{\textwidth}{!}{\includegraphics{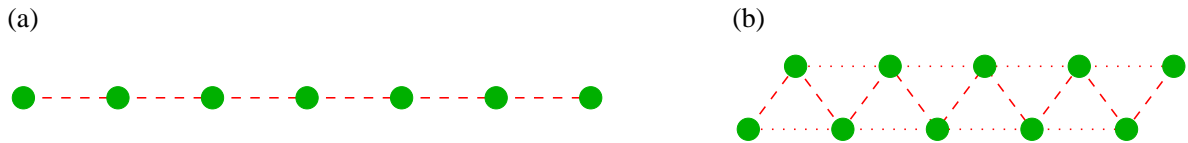}}
  \caption{\label{fig:crystalstructures} (a) A one-dimensional Wigner
    crystal formed in a quantum wire at low electron density.  (b) The
    zigzag Wigner crystal forms in a certain regime of densities when
    the electrons are confined to the wire by a shallow potential.}
\end{figure}

The electrons in quantum wires interact via repulsive Coulomb forces.
As a result of the long-range nature of the repulsion, at low density
the kinetic energy of the electrons is small compared to the
interactions.  To minimize their repulsion, electrons form a periodic
structure called the Wigner crystal \cite{Wigner}.  In one dimension
the long-range order in the Wigner crystal
(Figure~\ref{fig:crystalstructures}(a)) is smeared by quantum
fluctuations \cite{schulz}, and therefore the crystalline state can be
viewed as the strongly-interacting regime of the Luttinger liquid.
However, the presence of strong short-range order provides a clear
physical picture of the strongly interacting one-dimensional system
and enables one to develop a theoretical description of quantum wires
in the low-density regime.

In the Wigner crystal regime the electrons are strongly confined to
the vicinity of the lattice sites.  As a result the exchange of
electron spins is strongly suppressed, and only the nearest neighbor
spins are coupled to each other.  One can then think of the electron
spins forming a Heisenberg spin chain with a coupling constant $J$
much smaller than the Fermi energy $E_F$.  The presence of two very
different energy scales $E_F$ and $J$ for the charge and spin
excitations distinguishes the strongly interacting Wigner crystal
regime from a generic one-dimensional electron system with moderately
strong interactions.  In particular, the Luttinger liquid theory is
applicable to the Wigner crystal only at the lowest energies,
$\varepsilon\ll J$.  On the other hand, if any of the important energy
scales of the problem exceed $J$, the spin excitations can no longer
be treated as bosons, and the conventional Tomonaga-Luttinger picture
fails.  One of the most interesting examples of such behavior occurs
when the temperature $T$ is in the range $J\ll T\ll E_F$.  In this
case the charge excitations retain their bosonic properties consistent
with Luttinger liquid theory, whereas the correlations of electron
spins are completely destroyed by thermal fluctuations.  Such
one-dimensional systems are not limited to the Wigner crystal regime
and are generically referred to as spin-incoherent Luttinger liquids.
We argue in Sec.~\ref{sec:1D_crystal} that the coupling of spin and
charge excitations in this regime leads to a reduction of the
conductance of the quantum wire from $2e^2/h$ to $e^2/h$.  A number of
additional interesting properties of spin-incoherent Luttinger liquids
are discussed in a recent review \cite{fietereview}.

The electrons in a quantum wire are confined to one dimension by an
external potential.  In the common case of the potential created by
negatively charged gates placed on top of a two-dimensional electron
system, the confining potential can be rather shallow.  In this case
the strong repulsion between electrons can force them to move away
from the center of the wire, transforming the one-dimensional Wigner
crystal to a quasi-one-dimensional zigzag structure,
Figure~\ref{fig:crystalstructures}(b).  In the case of classical
electrons such a transition has been studied in
\cite{Chaplik,Hasse,Piacente}; we review this theory in
Sec.~\ref{sec:classical_zigzag}.  The zigzag Wigner crystal has rich
spin properties due to the fact that each electron can now be
surrounded by four neighbors with significant spin coupling.  Ring
exchange processes play an important role and may under certain
circumstances give rise to a spontaneous polarization of electron
spins.  The spin properties of the zigzag Wigner crystals are
discussed in Sec.~\ref{sec:spins}.

The transformation of a one-dimensional Wigner crystal to the zigzag
shape is a special case of a transition from a one-dimensional to a
quasi-one dimensional state of electrons in a quantum wire.  Another
such transition occurs in the case of noninteracting electrons when
the density is increased until population of the second subband of
electronic states in the confining potential begins.  These two
transitions seem to have rather different properties.  Indeed, in the
case of noninteracting electrons the population of the second subband
entails the emergence of a second acoustic excitation branch in the
system.  On the other hand, even though the zigzag crystal has two
rows, their relative motion is locked, and one expects to find only
one acoustic branch in this case.  It is therefore interesting to
explore how the number of acoustic excitation branches changes as the
interaction strength is tuned.  In the regime of strong interactions
this requires developing the quantum theory of the transition from a
one-dimensional to a zigzag Wigner crystal.  We discuss such a theory
in Sec.~\ref{sec:twomode}, where it is shown that quantum fluctuations
do not lead to the emergence of a second acoustic branch in the zigzag
crystal.  This feature of the Wigner crystal survives even at weak
interactions, with the second acoustic branch appearing only when the
interactions are completely turned off.

\section{One-dimensional crystal}
\label{sec:1D_crystal}

\subsection{Quantum wire at low electron density}
\label{sec:structure}

Electrons in a quantum wire repel each other with Coulomb forces.  To
characterize the strength of interactions, let us compare the typical
kinetic energy of an electron, which is of the order of the Fermi
energy $E_F\sim \hbar^2n^2/m$, with the typical interaction energy
$e^2n/\epsilon$. (Here $n$ is the electron density, $m$ is the
effective mass, and $\epsilon$ is the dielectric constant of the
medium.)  Clearly, the Coulomb repulsion dominates over the kinetic
energy in the low-density regime $na_B\ll 1$, where
$a_B=\hbar^2\epsilon/me^2$ is the Bohr's radius in the material.  Then
the ground state of the system is achieved by placing electrons at
well-defined points in the wire, separated from each other by the
distance $n^{-1}$, Figure~\ref{fig:crystalstructures}(a), thus
creating a Wigner crystal.  Because the kinetic energy of electrons is
small, the amplitude $\delta x$ of the zero-point fluctuations of
electrons near the sites of the Wigner lattice is much smaller than
the period of the crystal, $n\delta x\sim(na_B)^{1/4}\ll1$.

In experiment the quantum wire is usually surrounded by metal gates.
As a result, the Coulomb interactions between electrons are screened
at large distances by image charges in the gates.  For example, if the
gate is modeled by a metal plane at distance $d$ from the wire, the
interaction potential becomes
\begin{equation}
  \label{eq:screened_interaction}
  V(x)=\frac{e^2}{\epsilon}
  \left(\frac{1}{|x|}-\frac{1}{\sqrt{x^2+(2d)^2}}\right).
\end{equation}
At large distances this potential falls off as $V(x)\sim
2e^2d^2/\epsilon |x|^3$, much more rapidly than the original Coulomb
repulsion.  As a result, in the limit $n\to0$ the crystalline ordering
of electrons will be destroyed by quantum fluctuations.  Comparison of
the Fermi energy with the screened Coulomb repulsion
(\ref{eq:screened_interaction}) shows that the Wigner crystal exists
in the range of densities $a_B d^{-2}\ll n\ll a_B^{-1}$, provided that
the distance to the gate $d\gg a_B$.  In typical experiments with GaAs
quantum wire devices $a_B=10$nm and $d\gtrsim 100$nm; thus the Wigner
crystal state should persist until unrealistically low densities
$\sim10^{-3}{\rm nm}^{-1}$.

Similar to phonons in conventional crystals, the Wigner crystal
supports acoustic plasmon excitations---propagating waves of electron
density.  The speed of plasmons is given by\footnote{The result
  (\ref{eq:plasmon_speed}) was derived in \cite{Ruzin} for densities
  in the range $d^{-1}\ll n\ll a_B^{-1}$.  Extending their calculation
  to the density range $a_B/d^2\ll n\ll d^{-1}$, one finds
  $s=[24e^2n^3d^2\zeta(3)/\epsilon m]^{1/2}$.}
\begin{equation}
  \label{eq:plasmon_speed}
  s=\sqrt{\frac{2e^2n}{\epsilon m}\ln(8.0 nd)}.
\end{equation}
The Hamiltonian describing these low-energy excitations is easily
obtained by treating the Wigner crystal as a continuous medium.
Adding the kinetic energy and the potential energy of elastic
deformation, one obtains
\begin{equation}
  \label{eq:Wigner_Hamiltonian}
  H_\rho = \int \left[
    \frac{p^2}{2mn} + \frac12 mns^2 (\partial_x u)^2\right]dx,
\end{equation}
where $u(x)$ is the displacement of the medium at point $x$ from its
equilibrium position, and $p(x)$ is the momentum density.  In one
dimension the acoustic excitations destroy the long range order in the
crystal even at zero temperature, $\langle[u(x)-u(0)]^2\rangle\simeq
(\hbar/\pi mns)\ln nx$.\footnote{In the absence of the screening gate
  the plasmon speed $s$ diverges at small wavevectors, see
  (\ref{eq:plasmon_speed}) at $d\to\infty$.  Although this effect
  suppresses the quantum fluctuations, it is not sufficient to restore
  the long-range order \cite{schulz}.}

In the model of spinless electrons, Hamiltonian
(\ref{eq:Wigner_Hamiltonian}) accounts for all possible low-energy
excitations of the system.  However, in the presence of spins, there
are additional excitations not included in
(\ref{eq:Wigner_Hamiltonian}).  In the Wigner crystal regime the
electrons are localized near their lattice sites,
Figure~\ref{fig:crystalstructures}(a), and to a first approximation
the spins at different sites are not coupled.  The exchange coupling
of two spins at neighboring sites occurs via the process of two
electrons switching their places on the Wigner lattice.  When the
electrons approach each other, the strong Coulomb repulsion creates a
high potential barrier.  As a result, the exchange processes are very
weak, and only the coupling of the nearest neighbor spins needs to be
taken into account.  The Hamiltonian describing the spin excitations
takes the form
\begin{equation}
  \label{eq:spin-chain}
  H_\sigma = \sum_l J\, {\bm S}_{l}\cdot{\bm S}_{l+1},
\end{equation}
where ${\bm S}_{l}$ is the spin at site $l$.  As the exchange
processes involve tunneling through a high barrier, the exchange
constant is exponentially suppressed \cite{Hausler, Matveev,
  matveev-prb},
\begin{equation}
  \label{eq:J}
  J \propto \exp\left(-\frac{\eta}{\sqrt{na_B}}\right),
\end{equation}
where $\eta\approx 2.80$ \cite{KRM, Fogler1, Fogler2}, see also
Sec.~\ref{sec:numerics-1D}.  Taken together, equations
(\ref{eq:Wigner_Hamiltonian}) and (\ref{eq:spin-chain}) account for
all low-energy excitations of the one-dimensional Wigner crystal,
i.e., the Hamiltonian of the system can be represented as the sum
\begin{equation}
  \label{eq:separation}
  H = H_\rho + H_\sigma.
\end{equation}
Because of the absence of long-range order, one expects that in the
low-energy limit the Wigner crystal should be a special case of the
Luttinger liquid.  The latter is commonly described \cite{giamarchi}
by a Hamiltonian of the form (\ref{eq:separation}), with the charge
and spin Hamiltonians, $H_\rho$ and $H_\sigma$, given by
\begin{eqnarray}
  \label{eq:H_rho}
  \fl\qquad\qquad
  H_\rho=\int\frac{\hbar u_\rho}{2\pi}\left[\pi^2 K_\rho \Pi_\rho^2 
    +K_\rho^{-1}(\partial_x\phi_\rho)^2\right]dx,
  \\
  \fl\qquad\qquad
  \label{eq:H_sigma}
  H_\sigma=\int\frac{\hbar u_\sigma}{2\pi}
  \left[\pi^2 K_\sigma \Pi_\sigma^2 
    +K_\sigma^{-1}(\partial_x\phi_\sigma)^2\right]dx
  +\frac{2g_{1\perp}}{(2\pi\alpha)^2}\int 
  \cos\left[\sqrt8 \phi_\sigma(x)\right] dx.
\end{eqnarray}
Here the bosonic fields $\phi_{\rho,\sigma}$ and $\Pi_{\rho,\sigma}$
describe the charge ($\rho$) and spin ($\sigma$) excitations
propagating with velocities $u_{\rho,\sigma}$. They obey canonical
commutation relations $[\phi_\alpha(x), \Pi_{\alpha'}(y)] = \rmi
\delta_{\alpha\alpha'} \delta(x-y)$.  In the case of repulsive
interactions, the Luttinger liquid parameter $K_\rho$ is in the range
$0<K_\rho<1$.  The cosine term in (\ref{eq:H_sigma}) is marginally
irrelevant, i.e., the coupling constant $g_{1\perp}$ scales to zero
logarithmically at low energies.  At the same time, the parameter
$K_\sigma$ approaches unity as $K_\sigma = 1 + g_{1\perp}/2\pi
u_\sigma$.

Both Hamiltonians (\ref{eq:Wigner_Hamiltonian}) and (\ref{eq:H_rho})
describe propagation of elastic waves in the medium.  Their formal
equivalence is established \cite{matveev-prb} by identifying
\begin{equation}
  \label{eq:charge_identities}
  \phi_\rho(x) = \frac{\pi n}{\sqrt2}\,u(x) , 
  \quad 
  \Pi_\rho(x) = \frac{\sqrt2}{\pi n\hbar}\,p(x),
  \quad
  u_\rho=s,
  \quad
  K_\rho=\frac{\pi\hbar n}{2ms}.
\end{equation}
On the other hand, even though both Hamiltonians (\ref{eq:spin-chain})
and (\ref{eq:H_sigma}) describe spin excitations in the system, their
equivalence is not obvious.  Indeed, Hamiltonian (\ref{eq:spin-chain})
is expressed in terms of spin operators ${\bm S}_{l}$ of the
electrons, whereas its Luttinger-liquid analog (\ref{eq:H_sigma}) is
expressed in terms of the bosonic fields $\phi_\sigma$ and
$\Pi_\sigma$.  The connection is established via the well-known
procedure \cite{giamarchi} of bosonization of the Heisenberg spin
chain (\ref{eq:spin-chain}).  This procedure is applicable at energies
much smaller than the exchange constant $J$, and reduces Hamiltonian
(\ref{eq:spin-chain}) to the form (\ref{eq:H_sigma}), see
Ref.~\cite{matveev-prb}.  One therefore concludes that at low energies
the Wigner crystal can indeed be viewed as a Luttinger liquid.

It is important to point out, however, that the equivalence of the
Wigner crystal and Luttinger liquid holds only at very low energies,
$\varepsilon\ll J$.  Given the exponential dependence (\ref{eq:J}) of
the exchange constant on density, one can easily achieve a regime when
an important energy scale, such as the temperature, is larger than
$J$.  In this case the bosonization procedure leading to
(\ref{eq:H_sigma}) is inapplicable, and the form (\ref{eq:spin-chain})
should be used instead.  On the other hand, as long as temperature and
other relevant energy scales are smaller than the Fermi energy, the
charge excitations are bosonic and adequately described by either
Hamiltonian (\ref{eq:Wigner_Hamiltonian}) or (\ref{eq:H_rho}).

\subsection{Spin-charge separation in the one-dimensional Wigner
  crystal}
\label{sec:spin-charge_separation}

The Hamiltonian (\ref{eq:separation})-(\ref{eq:H_sigma}) of the
Luttinger liquid consists of two separate commuting contributions
associated with the charge and spin degrees of freedom.  Consequently,
the low-energy excitations of the system are charge and spin waves,
decoupled from each other, and propagating at different velocities
$u_\rho$ and $u_\sigma$.  The operator annihilating a (right-moving)
electron with spin $\gamma$ in this theory has the form
\begin{equation}
  \fl\qquad\qquad
  \psi_{R\gamma}(x)
  =\frac{\rme^{\rmi k_Fx}}{\sqrt{2\pi\alpha}}\,
  \exp\left\{\frac{\rmi}{\sqrt2}[\phi_\rho(x)-\theta_\rho(x)]\right\}
  \exp\left\{\pm\frac{\rmi}{\sqrt2}
    [\phi_\sigma(x)-\theta_\sigma(x)]\right\},
  \label{eq:electron_bosonization_right}
\end{equation}
in which the charge and spin contributions explicitly factorize.
(Here $\alpha$ is a short-distance cutoff, $k_F=\pi n/2$ is the Fermi
wavevector of the electrons, and the $+/-$ sign corresponds to
electron spin $\gamma=\uparrow,\downarrow$.)

The Hamiltonian of the Wigner crystal (\ref{eq:separation}) also
consists of two commuting contributions describing the charge and spin
degrees of freedom, with the main difference being the different form
(\ref{eq:spin-chain}) of $H_\sigma$.  However, the analogy with the
Luttinger liquid is not complete, as the electron annihilation
operator no longer factorizes \cite{mfl-prl,mfl-long},
\begin{equation}
  \psi_{R\gamma}(x)=\frac{\rme^{\rmi 2k_Fx}}{\sqrt{2\pi\alpha}}\, 
  \exp\left\{\frac{\rmi}{\sqrt2}[2\phi_\rho(x)-\theta_\rho(x)]\right\}
  Z_{l,\gamma}\Big|_{l=nx+\frac{\sqrt2}{\pi}\,\phi_\rho(x)}.
  \label{eq:our_annihilation_operator_bosonized}
\end{equation}
Here the operator $Z_{l,\gamma}$ acts upon any state of the spin chain
(\ref{eq:spin-chain}) and produces a state with one less spin by
removing spin $\gamma$ at site number $l$.  The form of the fermion
operator (\ref{eq:our_annihilation_operator_bosonized}) reflects the
fact that when an electron is removed from the Wigner crystal, one of
the sites of the spin chain (\ref{eq:spin-chain}) is also removed.  In
the absence of plasmon excitations, the sites are equidistant, and the
site at point $x$ has the number $l=nx$.  On the other hand, if
plasmons propagate though the crystal, the electrons shift by a
distance proportional to $\phi_\rho$, and the spin is removed from the
site $l = nx + \frac{\sqrt2}{\pi}\,\phi_\rho(x)$, see
(\ref{eq:our_annihilation_operator_bosonized}).  Thus the absence of
factorization of the charge and spin components of the fermion
operator (\ref{eq:our_annihilation_operator_bosonized}) reflects the
simple fact that the spins ${\bm S}_{l}$ in the spin chain
(\ref{eq:spin-chain}) are attached to the electrons.

The absence of spin-charge separation in the Hamiltonian of the Wigner
crystal manifests itself if the system is not uniform, such as in the
case of a quantum wire with a low electron density that depends on
position, $n=n(x)$.  Assuming that the variations of $n(x)$ occur at a
length scale much larger than the distance between electrons, one can
still bosonize the charge modes near every point in space, while
accounting for the $x$-dependence of the parameters $u_\rho$ and
$K_\rho$.  Thus one obtains
\begin{equation}
  \label{eq:H_rho_nonuniform}
  H_\rho=\int\frac{\hbar u_\rho(x)}{2\pi}\left\{\pi^2 K_\rho(x) \Pi_\rho^2 
    +[K_\rho(x)]^{-1}(\partial_x\phi_\rho)^2\right\}dx.
\end{equation}
The exchange constant $J$ in the Hamiltonian (\ref{eq:spin-chain}) of
the spin chain also acquires an $x$-dependence, as it clearly depends
on the electron density, see (\ref{eq:J}).  Thus the spin Hamiltonian
takes the form
\begin{equation}
  \label{eq:spin-chain_nonuniform}
  H_\sigma = \sum_l J\Big(l-\textstyle\frac{\sqrt2}{\pi}\,\phi_\rho(x_l)\Big)
  {\bm S}_{l}\cdot{\bm S}_{l+1},
\end{equation}
where $x_l$ is the initial position of the $l$-th electron.  The
appearance of the charge field $\phi_\rho$ in $H_\sigma$ again
accounts for the fact that the plasmons shift the site $l$ of the spin
chain from its initial position by $\frac{\sqrt2}{\pi}\,\phi_\rho$.
Therefore the two contributions $H_\rho$ and $H_\sigma$ to the
Hamiltonian of the Wigner crystal commute only in the uniform system,
when the exchange constant $J$ does not depend on position.

\subsection{Conductance of a Wigner-crystal wire}
\label{sec:conductance}

\begin{figure}[tb]
  \hspace{0.16\textwidth}
  \resizebox{0.5\textwidth}{!}{\includegraphics{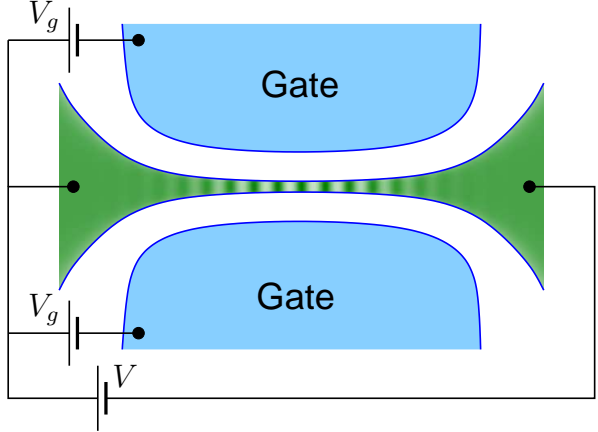}}
  \caption{\label{fig:wignerwire} A quantum wire formed by applying
    negative voltage to the gates placed on top of a two-dimensional
    electron system.  Electrons in the narrow channel between the
    gates are one-dimensional and their density is sufficiently low to
    achieve the Wigner crystal regime.  Away from the center of the
    wire the electron density increases, and even the short range
    ordering of electrons is destroyed by quantum fluctuations.}
\end{figure}

In experiment, the quantum wires are usually made by confining a
two-dimensional electron system to a one-dimensional channel.  One of
the most common techniques is to place two metal electrodes above a
GaAs heterostructure in which a two-dimensional electron system is
formed, Figure~\ref{fig:wignerwire}.  When a negative voltage is
applied to the gates, the resulting electrostatic potential repels the
electrons from the regions covered by the gates, but a narrow channel
of electrons between the gates may still remain.  The resulting
quantum wire connects two large regions of two-dimensional electrons,
which play the role of contacts to the wire.  If the gate voltage
$V_g$ is properly tuned, the electron density in the center of the
wire can be sufficiently low for a Wigner crystal to form.  On the
other hand, the gates do not affect the electron density and the
nature of the electron liquid in the two-dimensional leads.

The physics of interacting electrons in two or three dimensions is
very different from that of one-dimensional systems.  Although at
extremely low densities the electrons will form a Wigner crystal, this
does not happen in typical GaAs heterostructures.  Instead, the
electrons are believed to be in a conventional Fermi-liquid state with
quasiparticle excitations obeying Fermi statistics and carrying the
charge of a single electron.  In a one-dimensional system such a
situation may only occur in the absence of interactions, as otherwise
a Luttinger liquid state with bosonic excitations is formed.  In the
absence of interactions, however, the Fermi-liquid and
Luttinger-liquid pictures are equivalent.  Thus it is convenient to
model the quantum wire device by a one-dimensional model with
position-dependent interactions and electron density.  In the central
part of the system the density is small so that the interactions may
be effectively strong.  This region models the quantum wire.  As one
moves away from the central region, the density grows, the
interactions become small, and asymptotically at large distances the
electrons become noninteracting.  These two semi-infinite
noninteracting regions model the two-dimensional leads.

Such a model was used in \cite{maslov,ponomarenko,safi} to calculate
the conductance of a quantum wire described by the Luttinger liquid
model.  The Hamiltonian studied was essentially identical to
(\ref{eq:H_rho_nonuniform}), as the electrons were assumed to be
spinless and only charge modes needed to be accounted for.  It was
demonstrated that the dc conductance of the wire is not affected by
the interactions and remains quantized at $e^2/h$.  Let us illustrate
this result with a simple semiclassical calculation.

We start with the homogeneous wire, and for simplicity, instead of the
Hamiltonian (\ref{eq:H_rho}) we will use the equivalent form
(\ref{eq:Wigner_Hamiltonian}).  Unlike papers
\cite{maslov,ponomarenko,safi}, where a term was added to the
Hamiltonian in order to describe the bias voltage applied to the wire
at point $x=0$, we consider a setup in which the wire is connected to
a current source.  A small ac current with frequency $\omega$ can be
represented in terms of the velocity $\dot u$ of the elastic medium
and the electron density as
\begin{equation}
  \label{eq:current_velocity}
  ne\dot u|_{x=0} = I_0 \cos \omega t.
\end{equation}
This expression should be viewed as a time-dependent boundary
condition imposed on the elastic medium.  As a result the medium
begins to move periodically with frequency $\omega$, and plasmons
propagating into the infinite leads dissipate power $W=I_0^2 R/2$ from
the current source, where $R$ is the resistance of the system.  Let us
calculate $W$ in terms of the parameters of the elastic medium.  Since
the plasmons carry the energy of the oscillating medium in two
directions at speed $s$, we can express the dissipated power as
\begin{equation}
  \label{eq:power}
  W=2s\langle \mathcal E\rangle,
\end{equation}
where $\langle \mathcal E\rangle$ is the energy density of the system.
The latter consists of two contributions, the kinetic and potential
energies represented by the two terms in
(\ref{eq:Wigner_Hamiltonian}).  In a harmonic system the time-averaged
values of the kinetic and potential energies are equal, so we will
evaluate $\langle \mathcal E\rangle$ by doubling the kinetic energy,
\begin{equation}
  \label{eq:average_energy}
  \langle \mathcal E\rangle = mn\dot u^2 
  = \frac{m}{e^2n}\, I_0^2 \langle\cos^2\! \omega t\rangle
  = \frac{m}{2e^2n} I_0^2,
\end{equation}
where we expressed the velocity $\dot u$ in terms of the current using
(\ref{eq:current_velocity}).  Substituting this expression into
(\ref{eq:power}) and comparing the result with the Joule heat law
$W=I_0^2R/2$, we find the resistance
\begin{equation}
  \label{eq:impedance}
  R=\frac{2ms}{e^2n}=\frac{h}{e^2}\,\frac{s}{v_F},
\end{equation}
where we used the density $n=k_F/\pi$ for spinless electrons and
defined the Fermi velocity in the interacting system as $v_F=\hbar
k_F/m$.

In the noninteracting limit, where the Luttinger liquid theory
reproduces the low-energy properties of the Fermi gas, the plasmon
velocity $s=v_F$, and we recover the well-known result $R=h/e^2$.  The
model considered in \cite{maslov,ponomarenko,safi} was described by
the Hamiltonian (\ref{eq:H_rho_nonuniform}) of the inhomogeneous
Luttinger liquid, where the interactions are present only in a region
of finite size $L$, modeling the wire, and vanish at $x\to\pm\infty$.
It is easy to see that the above calculation of the resistance is
applicable to such a system as long as the low-frequency limit is
considered.  Indeed, at $\omega\to0$ the wavelength of the plasmons
$\sim s/\omega$ is much larger than $L$, so the emission of the
plasmons occurs in the noninteracting leads.  Thus we have recovered
the result \cite{maslov,ponomarenko,safi} for the conductance,
$G=e^2/h$.

Our simple calculation also enables us to interpret the absence of
corrections to the conductance due to electron-electron interactions
in a finite region of a one-dimensional system.  In the Luttinger
liquid theory the main effect of the interactions is to change the
compressibility of the electron system, thereby affecting the second
term in (\ref{eq:Wigner_Hamiltonian}).  In the dc limit the wavelength
of the plasmons is infinitely large, and thus the deformation
$\partial_x u$ within the finite-size interacting region is
negligible.  Thus the system behaves as a noninteracting one.

The above result for the spinless Luttinger liquid can be easily
generalized to the case of electrons with spin.  As we discussed in
Sec.~\ref{sec:spin-charge_separation}, within the Luttinger-liquid
approximation the charge and spin degrees of freedom are not coupled.
Thus the applied bias or electric current couples only to the charge
modes, and the above discussion can be repeated with the only
modification being the different relation $n=2k_F/\pi$ between the
density and the Fermi wavevector.  Substituting this expression
instead of $n=k_F/\pi$ in (\ref{eq:impedance}) we find the resistance
of the charge modes
\begin{equation}
  \label{eq:charge_resistance}
  R_\rho=\frac{h}{2e^2},
\end{equation}
and thus the expected doubling of the conductance, $G=2e^2/h$.

On the other hand, we saw in Sec.~\ref{sec:spin-charge_separation}
that in the inhomogeneous Wigner crystal there is no spin-charge
separation, i.e., the Hamiltonian (\ref{eq:spin-chain_nonuniform}) of
the spin excitations depends explicitly on the charge field
$\phi_\rho$.  One can therefore expect that the spin degrees of
freedom will affect conductance when the Wigner crystal is not
equivalent to the Luttinger liquid.  Indeed, we show below that the
spins have a significant effect on the electronic transport at
temperatures $T\gtrsim J$.

In treating a one-dimensional Wigner crystal attached to
noninteracting leads one has to overcome a fundamental problem caused
by the lack of quantitative theory for the crossover regions that
connect them, Figure~\ref{fig:wignerwire}.  In the case of spinless
electrons both the Wigner crystal and the leads can be viewed as
special cases of the Luttinger liquid, assuming that one is only
interested in the low-energy properties of the system.  Thus one can
use the model (\ref{eq:H_rho_nonuniform}) of the inhomogeneous
Luttinger liquid and obtain reliable results, provided that the exact
form of the $x$-dependences of the parameters is not important.  In
the presence of spins there is an additional complication caused by
the fact that the spin sector of a Wigner crystal is described by the
Hamiltonian of a Heisenberg spin chain (\ref{eq:spin-chain}) because
the spins are attached to well-localized electrons.  Such a
description is appropriate in neither the crossover region nor the
leads, where the short-range crystalline order is absent.  In our
further discussion we will nevertheless use the model of the
inhomogeneous spin chain (\ref{eq:spin-chain_nonuniform}) for the
whole system.  This model is justified if the temperature is small
compared to the Fermi energy in the center of the wire.  When one
moves away from the center, the density $n$ grows, and consequently
the exchange constant $J$ rapidly grows, see (\ref{eq:J}).  Even if in
the center of the system we had $J\ll T$, the crossover regime $J\sim
T$ will occur while the wire is still in the Wigner crystal regime, as
$J$ is still small compared to $E_F$.  Eventually, when one moves
sufficiently far from the center of the wire the exchange $J$ becomes
of order $E_F$, and the spin chain model is no longer appropriate.
However, since in those regions we have $J\gg T$, the Heisenberg model
(\ref{eq:spin-chain}) is equivalent to the spin sector
(\ref{eq:H_sigma}) of the Luttinger liquid theory appropriate for both
the crossover regions and the leads.  Thus, at $T\ll E_F$, one can
describe the spin properties of the system by the model
(\ref{eq:spin-chain_nonuniform}) of an inhomogeneous spin chain as
long as the exact shape of the dependence $J(l)$ does not affect the
results.

\begin{figure}[tb]
  \resizebox{\textwidth}{!}{\includegraphics{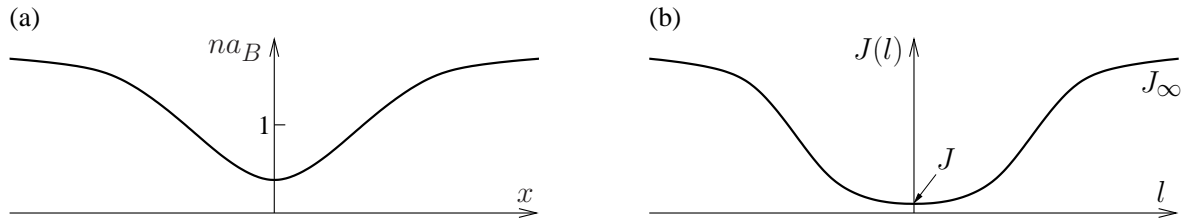}}
  \caption{\label{fig:density-exchange} (a) The electron density as a
    function of position has a minimum in the center of the wire
    ($x=0$), where $na_B\ll1$ and the Wigner crystal is formed.  In
    the lead regions, $na_B$ is assumed to be large such that the
    interactions can be neglected.  (b) The low density in the wire
    results in the exponential suppression (\ref{eq:J}) of the
    exchange constant $J$.  In the lead regions $J(l)$ saturates at
    $J_\infty\sim E_F$.}
\end{figure}

Formally the quantum wire will be described by the Hamiltonian
$H_\rho+H_\sigma$ given by (\ref{eq:H_rho_nonuniform}) and
(\ref{eq:spin-chain_nonuniform}).  The electron density has a minimum
at the center of the wire, resulting in an exponentially small
exchange constant $J$, Figure~\ref{fig:density-exchange}.  Far from
the center of the wire the exchange constant reaches the value
$J_\infty\sim E_F$.  Since $J$ depends on position, the spin
excitations are coupled to the charge excitations.  To find the
resulting correction to the conductance of the wire, it is convenient
to consider the setup of fixed current through the wire.  Given the
standard bosonization relation between $\partial_x\phi_\rho$ and the
electron density, by fixing the current $I$ at point $x=0$ one imposes
the boundary condition $\phi_\rho(0,t)=-(\pi/\sqrt2)q(t)$ on the
charge modes, where $q(t)$ is the charge transferred through the wire,
i.e., $I=e\dot q$.  As discussed above, at small frequencies $\omega$
the plasmon wavelength is very large, and electrons move in phase over
distances much longer that the length of the wire.  One can therefore
replace $\phi_\rho(x)$ by its value at $x=0$ everywhere within the
range where $J$ depends on position, and convert the Hamiltonian
(\ref{eq:spin-chain_nonuniform}) to the form
\begin{equation}
  \label{eq:spin-chain_q}
  H_\sigma = \sum_l J[l+q(t)]\,
  {\bm S}_{l}\cdot{\bm S}_{l+1}.
\end{equation}
The advantage of this form of $H_\sigma$ is that it now commutes with
$H_\rho$.  This does not mean that spin-charge separation is restored,
as the spin excitations are still affected by the electric current.

An immediate consequence of the commutativity of $H_\rho$ and
$H_\sigma$ is that the application of electric current through the
wire gives rise to independent excitation of the charge and spin
modes.  Assuming that the power dissipated in each channel is
quadratic in current, we conclude $W\equiv
I_0^2R/2=I_0^2(R_\rho+R_\sigma)/2$.  Thus the resistance of the wire
is a sum,
\begin{equation}
  \label{eq:series_resistance}
  R=R_\rho+R_\sigma,
\end{equation}
of two independent contributions due to the charge and spin
excitations.  Since we have already discussed the contribution
(\ref{eq:charge_resistance}) of the charge excitations, we now turn
our attention to $R_\sigma$.

The spin contribution to the resistance depends crucially on whether
the temperature is small or large compared to the value $J$ of the
exchange constant in the center of the wire, see
Figure~\ref{fig:density-exchange}.  At $T\ll J$ one can bosonize the
spin excitations, i.e., convert $H_\sigma$ to the form
(\ref{eq:H_sigma}) with position-dependent parameters.  Within this
approach, an attempt to account for the coupling to the charge modes
in (\ref{eq:spin-chain_nonuniform}) would result in corrections cubic
in the bosonic fields.  Such corrections are irrelevant perturbations,
which are usually neglected as their contribution vanishes at $T\to0$.
Thus one concludes that $R_\sigma=0$ in the limit $T/J\to0$.

\begin{figure}[tb]
  \hfill \resizebox{0.84\textwidth}{!}{\includegraphics{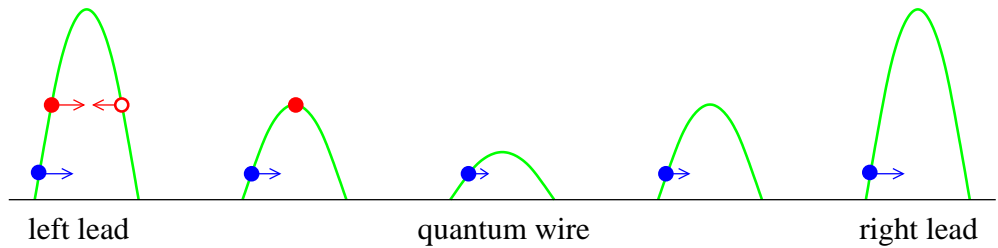}}
  \caption{\label{fig:spinons} Scattering of spinons at the quantum
    wire.  Spinons with energies below $\pi J/2$ (shown in blue) slow
    down in the wire, but continue to move forward to the opposite
    lead.  Spinons with energies above $\pi J/2$ (shown in red) stop
    before they reach the center of the wire and are scattered back.}
\end{figure}

The absence of dissipation in the spin channel at low temperature can
be interpreted as follows.  The low-energy excitations of a Heisenberg
spin chain are the so-called spinons \cite{faddeev} with spectrum
\begin{equation}
  \label{eq:spinons}
  \varepsilon(k)=\frac{\pi J}{2} \sin k,
\end{equation}
where the wavevector $k$ is defined in the interval $(0,\pi)$.  At low
temperature the state of the spin chain can be viewed as a dilute gas
of spinons.  Let us consider propagation of spinons in the spin chain
(\ref{eq:spin-chain_q}) with non-uniform $J$,
Figure~\ref{fig:density-exchange}(b), assuming for the moment
$q(t)=0$.  If the variation of $J(l)$ is very gradual, one can use the
spectrum (\ref{eq:spinons}) with $l$-dependent exchange $J$.  As a
spinon propagates through the wire, its energy is conserved, but its
momentum and velocity change because of the variation of $J$ along the
system.  Clearly, if the energy of a spinon is less than $\pi J/2$,
where $J$ is the smallest value of the exchange constant in the
system, Figure~\ref{fig:density-exchange}(b), it passes through the
wire without scattering.  Conversely, spinons with energies exceeding
$\pi J/2$ are backscattered, Figure~\ref{fig:spinons}.

At $q(t)\neq0$ the dependence $J(l)$ shown in
Figure~\ref{fig:density-exchange}(b) is not static, but rather
oscillates in position with respect to the spin chain.  (More
physically, the ac current moves the Wigner crystal with respect to
the quantum wire, causing the time dependence of the exchange
constants in (\ref{eq:spin-chain_q}).)  The spinons passing through
the wire without scattering are not affected by this oscillation.  On
the other hand, the spinons with energies $\varepsilon>\pi J/2$ are
reflected by a \emph{moving} scatterer.  Such processes do change the
energy of the spinons, and eventually lead to dissipation.  At low
temperature $T\ll J$ the density of such (thermally-activated) spinons
is very low, and one expects only an exponentially small resistance in
this regime,
\begin{equation}
  \label{eq:R_sigma_low_T}
  R_\sigma\propto\exp\left(-\frac{\pi J}{2T}\right), \qquad T\ll J.
\end{equation}
It is worth mentioning that the resistance (\ref{eq:R_sigma_low_T}) is
caused by excitations with energies of the order of the spinon
bandwidth $J$.  Such a correction cannot in principle be obtained by
the bosonization procedure, which is accurate only at energies much
smaller than $J$.

The expression (\ref{eq:R_sigma_low_T}) implies that the resistance
$R_\sigma$ grows with temperature.  At $T\gg J$ one expects this
growth to saturate.  Indeed, in this limit one can assume that $J=0$
in the center of the wire, i.e., the propagation of spin excitations
through the wire is no longer possible.  On the other hand, in the
leads one still has $T\ll J_\infty\sim E_F$, and the picture of a
dilute spinon gas still applies.  Every spinon moving toward the wire
is reflected back, resulting in a finite dissipation that no longer
depends on $J$.

Unfortunately, one cannot easily develop the theory of scattering of
spinons in this regime, as such processes occur in the region where
$J(l)\sim T$, and the spinon gas is no longer dilute.  One can,
however, conjecture that the dissipation resulting from all the spin
excitations being reflected by the wire is universal in the sense that
it does not depend on the exact nature of the scatterer.  Thus if one
can solve another problem where all the spin excitations in a
one-dimensional system are reflected by a moving scatterer, the result
for $R_\sigma$ should be the same.  The simplest example of such a
problem is obtained in the same Wigner-crystal setup in the presence
of a magnetic field $B$ sufficient to polarize electrons in the center
of the wire, $ T, J\ll \mu_B B\ll E_F$, where $\mu_B$ is the Bohr
magneton.  Then only the electrons with spin directed along the field
propagate through the wire whereas the electrons with opposite spin
are confined to the leads. This problem can be easily solved in the
framework of the bosonization approach \cite{matveev-prb}, resulting
in
\begin{equation}
  \label{eq:R_sigma_high_T}
  R_\sigma = \frac{h}{2e^2}, \qquad T\gg J.
\end{equation}
The result is easily understood by noticing that in combination with
(\ref{eq:series_resistance}) and (\ref{eq:charge_resistance}) one
finds the conductance $G=e^2/h$ which is the expected result for the
conductance of a spin-polarized wire, where only one type of charge
carriers participates in conduction.  By our conjecture, the same
reduction of conductance from $2e^2/h$ to $e^2/h$ occurs in the
absence of the field, provided $J\ll T$, because in both cases all the
spin excitations are reflected by the wire, resulting in the same
dissipation.  This conclusion is consistent with some of the
measurements of the conductance of quantum wires a low density
\cite{Thomas_2, Crook, Kane, Reilly_1}, showing a small plateau at
$G=e^2/h$.

\section{Classical transition to the zigzag structure}
\label{sec:classical_zigzag}

In section \ref{sec:1D_crystal}, we discussed the physics of a purely
one-dimensional crystal.  Experimentally, however, quantum wires are
created by confining three-dimensional electrons to a narrow channel
by an external confining potential.  The electron system in the wire
can be viewed as one-dimensional as long as the typical energy of the
transverse motion is large compared with all other important energy
scales; otherwise, deviations from one-dimensionality arise.  The
remainder of this review addresses the resulting quasi-one-dimensional
physics, starting with the classical transition from a one-dimensional
to a quasi-one-dimensional Wigner crystal that was studied in
Refs.~\cite{Chaplik,Hasse,Piacente}.

To be specific, we consider here a confining potential that mimics the
experimental situation. In a typical setup the confining potential in
one direction, say the $z$-direction, is provided by the band bending
at the interface of two semiconductors with different band structure
(typically GaAs and AlGaAs).  This provides a very tight confinement
and, correspondingly, the energy scales for transverse excitations are
large.  Therefore, at low energies, the possibility of electron motion
in the $z$-direction may be neglected.  By contrast, confinement in
the $y$-direction is provided by nearby metallic gates which create a
relatively shallow confining potential.  Deviations from
one-dimensionality arise due to lateral displacements in this shallow
potential which may be assumed parabolic:
\begin{eqnarray}
  V_{\rm conf}\eq\frac12m\Omega^2\sum_i y_i^2,
\end{eqnarray}
where $\Omega$ is the frequency of harmonic oscillations in the
confining potential, and $y_i$ is the transverse coordinate of the
electron at site $i$.

As the electron density $n$ grows, so does the typical energy $V_{\rm
  int}\sim (e^2/\epsilon)n$ of the Coulomb interaction between
electrons.  Eventually, it becomes energetically favorable for
electrons to move away from the axis of the wire.  This happens when
the distance between particles is of the order of the length scale
\begin{eqnarray}
  r_0=\sqrt[3]{\frac{2e^2}{\epsilon m\Omega^2}},
\end{eqnarray}
defined by the condition that the confinement and the Coulomb
repulsion, $V_{\rm conf}(r_0)=\frac12m\Omega^2r_0^2$ and $V_{\rm
  int}(r_0)=e^2/\epsilon r_0$, are equal~\cite{Piacente}.

The quasi-one-dimensional arrangement that maximizes the distance
between electrons---and consequently minimizes the Coulomb interaction
energy $V_{\rm int}=(e^2/\epsilon)\sum_{i<j}|{\bf r}_i-{\bf
  r}_j|^{-1}$---at a given cost of confining potential energy is a
zigzag structure, see Figure \ref{fig:crystalstructures}(b).  The
exact shape of the zigzag crystal can be found by minimizing its
energy per particle
\begin{eqnarray}
  E=\frac{e^2}{\epsilon r_0}
  \left\{\frac\nu2\sum_{l=1}^{\infty}\left(\frac1{l} +
      \frac1{\sqrt{(l-\frac12)^2 + \frac{\nu^2w^2}{4r_0^2}}}\right) +
    \frac{w^2}{4r_0^2}\right\} 
\end{eqnarray}
with respect to the distance $w$ between the two rows of the zigzag
crystal.  Here $\nu=nr_0$ is the dimensionless density, the first two
terms account for the interactions between electrons within the same
row and in different rows of the zigzag structure, respectively, and
the last term stems from the confining potential.

One finds that the distance between rows is given by the solution of
the equation
\begin{eqnarray}
  \left(
    \frac{\nu^3}{4}\sum_{l=1}^\infty
    \frac1{\bigg[(l-\frac12)^2 + \frac{\nu^2w^2}{4r_0^2}\bigg]^{3/2}}-1
  \right)w=0.
\end{eqnarray}
Below the critical density~\cite{Chaplik,Hasse}
\begin{eqnarray}
  \nu_c=\sqrt[3]{\frac4{7\zeta(3)}}\approx 0.780,
\end{eqnarray}
the only solution is $w=0$ and, therefore, the crystal is
one-dimensional.  At densities, $\nu>\nu_c$, a lower-energy solution
with $w\neq0$ appears, and the zigzag structure is formed.  The
distance between the two rows of the zigzag crystal grows with
density.  In particular, just above the transition point $\nu_c$, the
distance between rows behaves as
$w=r_0[\sqrt{{24}/{93\zeta(5)}}/\nu_c^{2}]\sqrt{\delta \nu}$, where
$\delta\nu=\nu-\nu_c$. Upon further increasing the density, the zigzag
crystal eventually becomes unstable at $\nu \approx 1.75$.  At larger
densities, $\nu>1.75$, structures with more than two rows are
energetically favorable~\cite{Piacente}.

Such a classical description of the system is valid only in the limit
where the distance between electrons is much larger than the Bohr's
radius, $n^{-1}\gg a_B$. As the zigzag regime corresponds to distances
between electrons of order $r_0$, it can only be achieved if $r_0$ is
sufficiently large, $r_0\gg a_B$. This motivates the introduction of a
density-independent parameter
\begin{eqnarray}
  r_\Omega=\frac{r_0}{a_B},
  \label{eq:r-Omega}
\end{eqnarray}
which characterizes the strength of Coulomb interactions with respect
to the confining potential.  If $r_\Omega\ll1$, as the electron
density grows, the interactions become weak at $n\sim a_B^{-1}\ll
r_0^{-1}$.  As a result, the one-dimensional Wigner crystal melts by
quantum fluctuations before the zigzag regime is reached.  By
contrast, if $r_\Omega\gg1$, interactions are still strong
($na_B\ll1$) at densities $n\sim r_0^{-1}$, and the classical
description of the transition to the zigzag regime is applicable.  As
$r_\Omega\propto\Omega^{-2/3}$, the strongly interacting case
therefore requires a shallow confining potential. Note that the
condition $r_\Omega\gg1$ can be rewritten as $W\gg a_B$, where
$W=\sqrt{\hbar/ m \Omega}$ is the (quantum) width of the wire.

\section{Spin properties of zigzag Wigner crystals}
\label{sec:spins}

In a Wigner crystal electrons are localized near their lattice
positions due to the mutual Coulomb repulsion. The potential landscape
thus created is such that any deviation from these lattice positions
incurs an increase in Coulomb energy. In particular, the exchange
processes which give rise to spin-spin interactions require tunneling
of electrons through the Coulomb barrier that separates them. As
pointed out in section \ref{sec:structure}, the resulting spin
couplings in a one-dimensional crystal are fairly simple: as the
tunneling amplitude decays exponentially with distance, only nearest
neighbor exchange processes have to be taken into account. Thus, the
spin degrees of freedom of a one-dimensional Wigner crystal are
described by an antiferromagnetic Heisenberg chain
(\ref{eq:spin-chain}) with nearest neighbor exchange energy $J$ whose
properties were discussed in section \ref{sec:1D_crystal}.

In a zigzag chain, spin couplings become more interesting. Close to
the zigzag transition, the nearest neighbor exchange is dominant as in
the one-dimensional case. However, as the zigzag structure becomes
more pronounced each electron is surrounded by four close neighbors
rather than only two as in the one-dimensional crystal, and,
therefore, the next-nearest neighbor couplings can no longer be
neglected.  Instead of one coupling constant, one needs to take into
account a nearest neighbor exchange constant $J_1$ and a next-nearest
neighbor exchange constant $J_2$. Both couplings are antiferromagnetic
and, therefore, compete with each other. If $J_2$ is large enough
($J_2\gtrsim 0.24..\,J_1$ \cite{Haldane,Okamoto,eggert}), the
antiferromagnetic ground state gives way to a dimer phase
characterized by a non-vanishing order parameter $D\propto\langle({\bm
  S}_{2i+1}-{\bm S}_{2i-1})\cdot{\bm S}_{2i}\rangle$ and a resulting
spin gap. The dimer structure is particularly simple on the so-called
Majumdar-Ghosh \cite{majumdar-ghosh1,majumdar-ghosh2} line
$J_2=0.5J_1$, where the dimers are just nearest neighbor singlets. The
magnitude of the spin gap is a non-monotonic function of the ratio
$J_2/J_1$: it reaches its maximum close to the Majumdar-Ghosh line and
becomes exponentially small at $J_2\gg J_1$.

It turns out, however, that these two-particle exchanges are not
sufficient to describe the spin physics of the zigzag crystal. In
addition, ring exchanges, i.e., cyclic exchanges of $n\geq3$
particles, have to be taken into account. Defining exchange constants
in such a way that they are all positive, the Hamiltonian of the
system then reads
\begin{eqnarray}
  \label{eq:ring-Hamiltonian}
  H_{\rm
    ring}&=&\frac12\sum_l\Big(J_1P_{l\,l\!+\!1} + J_2P_{l\,l\!+\!2} -
  J_3(P_{l\,l\!+\!1\,l\!+\!2}+P_{l\!+\!2\,l\!+\!1\,l})\nn\\ 
  &&+J_4(P_{l\,l\!+\!1\,l\!+\!3\,l\!+\!2}+P_{l\!+\!2\,l\!+\!3\,l\!+\!1\,l})-\dots\Big),
\end{eqnarray}
where $P_{ik}$ is a permutation operator and $P_{i_1\dots
  i_N}=P_{i_1i_2}P_{i_2i_3}\dots P_{i_Ni_1}$.  Here we still label
particles according to their position along the wire axis only: thus,
nearest neighbors are particles in opposite rows whereas next-nearest
neighbors are the closest particles within the same row.  Note that
for densities in the range $1.45<\nu< 1.75$ the lateral displacement
$w$ is so large that the distance between nearest neighbors becomes
larger than the distance between next-nearest neighbors.

Ring exchanges are interesting because they might stabilize a
ferromagnetic ground state. While exchanges involving even numbers of
particles favor a spin-zero ground state, exchanges involving odd
numbers of particles favor a ferromagnetic arrangement of spins
\cite{Thouless}.  Thus, the simplest ring exchange process that could
lead to a polarized ground state is the three-particle exchange.  In
fact, ring exchanges have been extensively studied in two-dimensional
Wigner crystals \cite{Roger,Katano,Voelker,ceperley}. In that case the
three-particle ring exchange dominates in the low-density limit which
implies a ferromagnetic ground state of the strongly interacting
Wigner crystal in two dimensions.\footnote{The ferromagnetic state is
  predicted to occur only at extremely low densities characterized by
  a value of $r_s>175$ \cite{ceperley}, where $r_s$ is the ratio of
  the Coulomb interaction energy to the Fermi energy.} To find out
whether the physics of the zigzag Wigner crystal is similar, one needs
to compute the exchange constants for nearest neighbor, next-nearest
neighbor, and the various ring exchanges.

\subsection{Computation of exchange constants}

To introduce the method, we start by discussing the one-dimensional
case where the only non-negligible exchange is the nearest neighbor
exchange.

\subsubsection{Exchange constants for the one-dimensional Wigner
  crystal}
\label{sec:numerics-1D}

The nearest neighbor exchange constant $J$ can be determined by
computing the tunneling probability of two electrons through the
Coulomb barrier that separates them. If the barrier is sufficiently
high and, therefore, tunneling is weak, one may use the semiclassical
instanton approximation.  This corresponds to finding the classical
exchange path in the inverted potential by minimizing the
imaginary-time action.

It is convenient to rewrite the action in dimensionless form by
rescaling length in units of $1/n$ and time in units of
$\sqrt{\epsilon m/e^2n^3}$.  The action of the system is then given as
\begin{eqnarray}
  \label{eq:eta-1D}
  \fl S_{1{\rm D}}=\frac\hbar{\sqrt{na_B}}\eta_{1{\rm
      D}}, \quad {\rm where} \quad \eta_{1{\rm D}}[\{x_j(\tau)\}]\eq
  \int \rmd\tau \left[\sum_{j}
    \frac{\dot{x}_j^2}{2}
    +\sum_{j<i}\frac{1}{|x_j-x_i|} \right].
\end{eqnarray}
As a first approximation one may fix the positions of all particles
except the two that participate in the exchange process, say $j=1$ and
$j=2$.  Symmetry fixes the center of mass coordinate of the exchanging
electrons and, therefore, the minimization has to be done only with
respect to the relative coordinate $x=x_2-x_1$. The tunneling lifts
the ground state degeneracy present due to inversion symmetry
$x\to-x$, and the exchange energy can be identified with the resulting
level splitting.

The instanton approximation yields the exchange constant $J$ in the
form (\ref{eq:J}), where $\eta$ is the dimensionless classical action
obtained from the minimization procedure.  One finds
$\eta\approx2.817$~\cite{KRM}.  At low densities, $na_B\ll1$, the
exponent is large leading to exponential suppression of $J$, and thus
the prefactor omitted in (\ref{eq:J}) is of secondary importance.

Fixing the positions of all particles except the two participating in
the exchange process is a somewhat crude approximation.  Neighboring
electrons see a modified potential due the motion of the exchanging
particles and, therefore, experience a force that displaces them from
their equilibrium positions.  A better estimate for $\eta$ can be
obtained by including these mobile ``spectator'' particles in the
minimization.  By allowing spectators to move during the exchange
process, one expects to find a reduced value for $\eta$ because more
variables are varied in the minimization procedure.  It turns out,
however, that the effect is very small.  As more spectators are added,
$\eta$ approaches the asymptotic value
$\eta\approx2.798$~\cite{KRM,Fogler1,Fogler2}, i.e., the result
changes by less than $1\%$.

\subsubsection{Exchange constants for the zigzag Wigner crystal}
\label{sec:numerics-zigzag}

\begin{figure}[tb]
  \hfill \resizebox{0.84\textwidth}{!}{\includegraphics{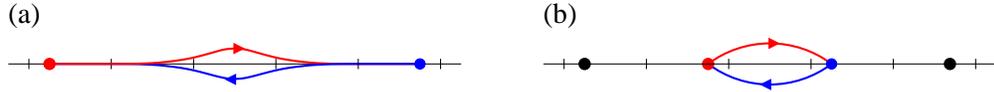}}
  \caption{\label{fig:1Dinstanton} Sketch of typical exchange paths
    for (a) $\nu\ll\nu_c$ and (b) $\nu\lesssim\nu_c$. the size of the
    loop where electrons move away from the axis of the wire is
    determined by the length scale $r_0$.}
\end{figure}

In the presence of a confining potential, the motion of the exchanging
electrons is no longer restricted to one dimension, i.e., the position
of an electron is now given by a two-dimensional vector ${\bf
  r}_j=(x_j,y_j)$. In particular, if the wire width $W$ is larger than
the Bohr's radius $a_B$ or, equivalently, the interaction parameter
introduced in Eq.~(\ref{eq:r-Omega}) is large, $r_\Omega\gg1$,
electrons can make use of the transverse direction to go ``around''
rather than ``through'' each other during the exchange process. This
reduces the Coulomb barrier and, therefore, increases the tunneling
probability. The characteristic length scale of the transverse
displacement is given by the length $r_0$, introduced in section
\ref{sec:classical_zigzag}.  Typical trajectories for the
one-dimensional crystal are shown in Figure \ref{fig:1Dinstanton} for
low and moderate densities, $\nu\ll\nu_c$ and $\nu\lesssim \nu_c$,
respectively. At low densities, $\nu\ll\nu_c$, the exchange part
follows the bottom of the confining potential until electrons come
within a distance of order $r_0$ of each other. Thus, only a small
part of the exchange path explores the transverse direction, leading
to a relatively small correction to the tunneling action $S_{1\rm D}$.
The results of section \ref{sec:numerics-1D} are recovered in the
limit $\nu\to0$. As one approaches the transition to the zigzag
crystal, the exchange trajectories become more and more
two-dimensional and consequently the exchange couplings are modified
significantly. Finally, at $\nu>\nu_c$, also the equilibrium positions
of the particles are displaced in the $y$-direction.

The exchange constants for the zigzag Wigner crystal can be obtained
in the same way as for the one-dimensional Wigner crystal~\cite{kmm}.
However, by contrast to the one-dimensional case, the structure of the
zigzag crystal changes as a function of density.  As a consequence the
rescaling of lengths and times used in the one-dimensional case is not
appropriate here.  A dimensionless action in a transverse confining
potential is conveniently defined using the interaction parameter
$r_\Omega$. Namely
\begin{eqnarray}
  \label{eq:eta-zigzag}
  \fl S_{\rm 2D}=\hbar\sqrt{r_\Omega}\; \eta_{\rm 2D}, \quad {\rm where}
  \quad \eta_{\rm 2D}[\{{\bf r}_j(\tau)\}]\eq\int
  \rmd \tau \left[\sum_{j}
    \left(\frac{\dot{{\bf r}}_j^2}{2}+y_j^2\right)
    +\sum_{j<i}\frac{1}{|{\bf r}_j-{\bf r}_i|} \right].
\end{eqnarray}
Here lengths have been rescaled in units of $r_0$ whereas times has
been rescaled in units of $\sqrt2/\Omega$. Furthermore, comparing
(\ref{eq:eta-1D}) and (\ref{eq:eta-zigzag}), the differences are the
one-dimensional vs two-dimensional coordinates and the additional term
due to the confining potential in (\ref{eq:eta-zigzag}).

As a result the exchange constants take the form
\begin{eqnarray}
  J_l\eq J_l^*\exp\left(-\eta_l\sqrt{r_\Omega}\right),
\end{eqnarray}
where $J_1$ is the nearest-neighbor exchange constant, $J_2$ is the
next-nearest neighbor exchange constant, and $J_l$ for $l\geq3$ is the
exchange constant corresponding to the $l$-particle ring exchange. The
exponents $\eta_l$ are obtained by minimizing the dimensionless action
$\eta_{\rm 2D}[\{{\bf r}_j(\tau)\}]$ for a given exchange process.
Whereas in the strictly one-dimensional case $\eta$ was just a number,
now the electron configuration changes as a function of density and,
therefore, the exponents $\eta_l$ depend on density, too.

\begin{figure}[tb]
  \hfill \resizebox{\textwidth}{!}{\includegraphics{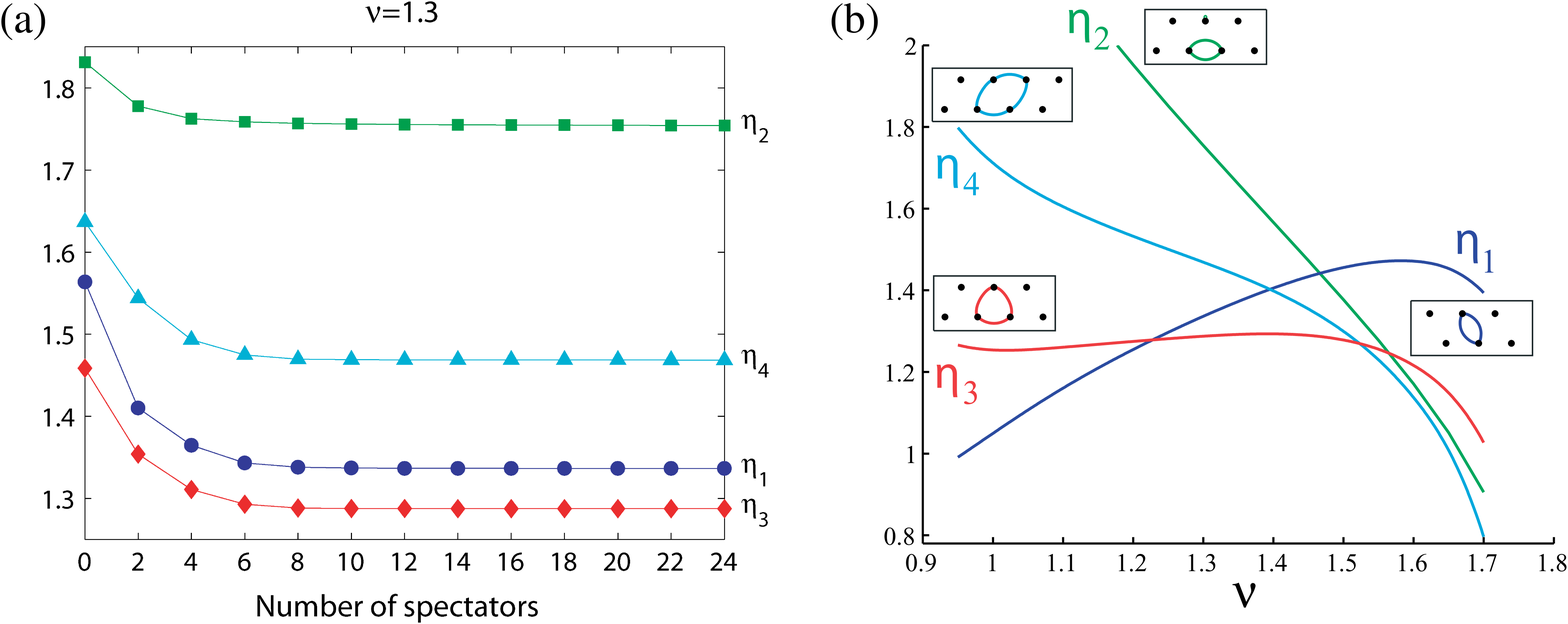}}
  \caption{\label{fig:eta_l} (a) Dependence of the exponents $\eta_l$
    on the number of spectators included in the
    calculation~\cite{klironomos-unpub}. Results are shown for
    $\nu=1.3$.  (b) Exponents $\eta_l$ for the nearest neighbor,
    next-nearest neighbor, three-particle ring, and four-particle ring
    exchange as a function of dimensionless density
    $\nu$~\cite{kmhm}.}
\end{figure}

Note that while in the one-dimensional case the inclusion of
spectators had little effect on the results, here the spectators turn
out to be much more important~\cite{kmm}. Figure \ref{fig:eta_l}(a)
shows the change of the exponents $\eta_l$ as spectators are included.
As one can see, the first few spectators modify the results
significantly. However, the results converge rapidly as more and more
spectators are added. Thus, the spin couplings are generated by
processes that involve the motion of a small number of close-by
electrons. Therefore, these couplings should not be affected by
deviations from the perfect crystalline order at large distances,
caused by quantum fluctuations.

Figure \ref{fig:eta_l}(b) shows the exponents $\eta_l$ as a function
of the dimensionless density $\nu$~\cite{kmm,kmhm}. Ring exchanges
with more than four particles are not included as they are negligibly
small at all densities. At small $\nu\lesssim1.2$, the crystal
geometry is still close to one-dimensional. In that regime $\eta_1$ is
the smallest exponent and therefore, as expected, the nearest-neighbor
exchange $J_1$ dominates.  However, as density increases and the
distances between nearest neighbors and next-nearest neighbors become
comparable, in the regime $1.2\lesssim\nu\lesssim1.5$ the
three-particle ring exchange constant $J_3$ becomes largest. Finally,
at even higher densities $\nu\gtrsim1.5$ the four-particle ring
exchange is dominant (until the zigzag crystal gives way to structures
with more than two rows at $\nu\approx 1.75$). In the next section the
ground states generated by these spin couplings will be discussed.

\subsection{Spin phases of the zigzag Wigner crystal}
\label{sec:zigzag-phases}

\begin{figure}[tb]
  \hfill \resizebox{\textwidth}{!}{\includegraphics{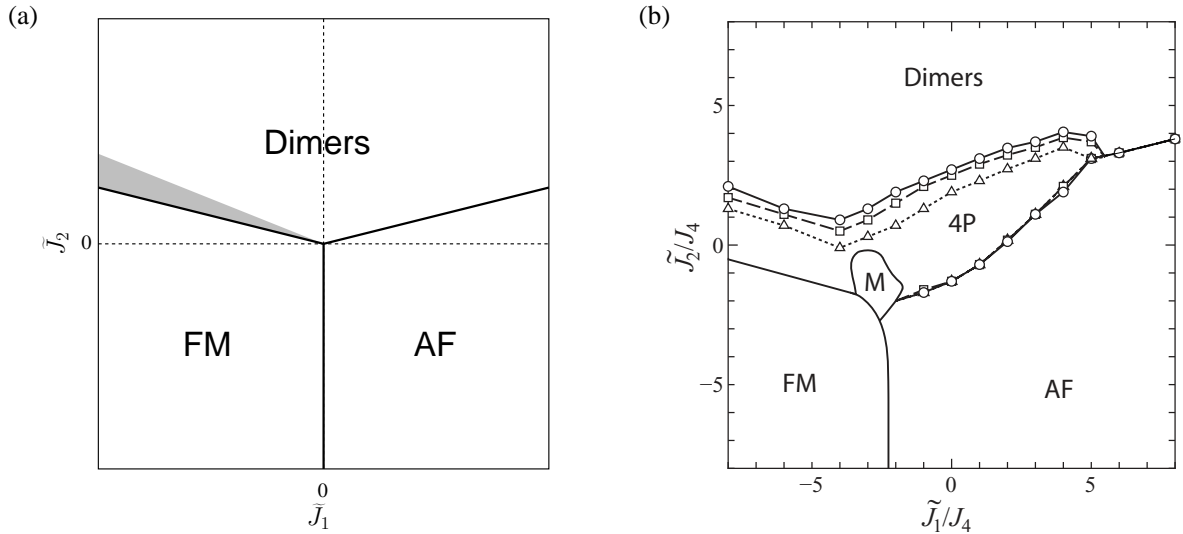}}
  \caption{\label{fig:J4} (a) Phase diagram of the Heisenberg spin
    chain with nearest neighbor coupling $\widetilde{J}_1$ and
    next-nearest neighbor coupling $\widetilde{J}_2$. (b) Preliminary
    phase diagram of the zigzag spin chain including four-particle
    ring exchange $J_4$, obtained by exact numerical diagonalization
    of finite-size chains \cite{kmhm}.  When $J_4$ is large, novel
    phases appear.  (Triangles, squares, and circles correspond to the
    boundaries obtained for $N$=16, 20, and 24 sites, respectively.)}
\end{figure}

In order to extract the spin properties of the ground state, it is
convenient to rewrite Hamiltonian (\ref{eq:ring-Hamiltonian}) in terms
of spin operators using the identity $P_{ik}=\frac12+2\, {\bm
  S}_i\cdot{\bm S}_k$.  In the absence of ring exchanges the system is
described as a Heisenberg spin chain with nearest neighbor and
next-nearest neighbor coupling,
\begin{eqnarray}
  \label{eq:J1J2}
  H_{12}\eq\sum_l \left(J_1\, {\bm S}_{l}\cdot{\bm S}_{l+1}+J_2\, {\bm
      S}_{l}\cdot{\bm S}_{l+2}\right). 
\end{eqnarray}
As discussed at the beginning of this section, depending on the ratio
of $J_1$ and $J_2$, one finds an antiferromagnetic and a dimer phase.
The contribution of the three-particle ring exchange is
\begin{eqnarray}
  H_3\eq-J_3\sum_l\left(2{\bm S}_{l}\cdot{\bm S}_{l+1} + {\bm
      S}_{l}\cdot{\bm S}_{l+2}\right). 
\end{eqnarray}
Thus, no new terms are generated---the Hamiltonian retains the same
form (\ref{eq:J1J2}), albeit with modified coupling constants
\begin{eqnarray}
  \widetilde{J_1}=J_1-2J_3,\qquad\qquad\widetilde{J_2}=J_2-J_3.
\end{eqnarray}
The important consequence is that the new coupling constants
$\widetilde{J_1}$ and $\widetilde{J_2}$ may now be either positive or
negative, corresponding to antiferromagnetic or ferromagnetic
interactions, respectively. The phase diagram of a Heisenberg spin
chain with both antiferromagnetic and ferromagnetic nearest and
next-nearest neighbor couplings has been widely studied in the
literature
\cite{Haldane,Okamoto,eggert,majumdar-ghosh1,majumdar-ghosh2,White,Hamada,Tonegawa,Chubukov,Allen,Itoi}.
In addition to the antiferromagnetic and dimer phases existing for
positive couplings, a ferromagnetic phase appears. The phase diagram
is shown in Figure \ref{fig:J4}(a).

This phase diagram is sufficient to determine the ground state of the
strongly-interacting zigzag Wigner crystal at low and intermediate
densities. At low densities, the system is in the antiferromagnetic
phase ($\widetilde{J_1}>0$, $|\widetilde{J_2}|\ll\widetilde{J_1}$). At
intermediate densities, the three particle ring exchange dominates.
As a result both coupling constants become negative,
$\widetilde{J_1},\widetilde{J_2}<0$, and therefore the system is in
the ferromagnetic phase.  The spontaneous spin polarization suggested
as a possible explanation of the 0.7 anomaly can, thus, occur in
strongly interacting quantum wires, if deviations from
one-dimensionality are taken into account.

At higher densities, the situation becomes more complicated. While the
three-particle ring exchange only modifies the nearest neighbor and
next-nearest neighbor exchange constants, the four-particle ring
exchange generates new terms in the Hamiltonian, namely a
next-next-nearest neighbor exchange and, more importantly, four-spin
couplings. The corresponding spin Hamiltonian reads
\begin{eqnarray}
  \fl H_4\eq J_4\sum_l\Big(\sum_{n=1}^3\frac{4-n}{2} {\bm S}_l\cdot{\bm
    S}_{l+n}\\
  \fl && +2\left[({\bm S}_l\cdot{\bm S}_{l+1})({\bm S}_{l+2}
    \cdot{\bm S}_{l+3})+({\bm S}_l\cdot{\bm S}_{l+2})({\bm S}_{l+1}\cdot {\bm
      S}_{l+3})-({\bm S}_l\cdot{\bm S}_{l+3})({\bm S}_{l+1} \cdot{\bm
      S}_{l+2})\right]\Big). \nn
\end{eqnarray}
The phase diagram in the presence of these couplings is not yet fully
understood.  First results were obtained using exact diagonalization
of short chains with up to $N=24$ spins \cite{kmhm}.  If
$J_4\ll|\widetilde{J_1}|,|\widetilde{J_2}|$, the same phases as in the
Heisenberg spin chain without four-particle ring exchange appear as
can be seen in Figure \ref{fig:J4}(b). However, as $J_4$ becomes of
the same order as the other coupling constants new phases appear.  The
simplest one to identify is a partially polarized phase (labeled `M'
in Figure \ref{fig:J4}(b)) adjacent to the ferromagnetic phase. While
this phase seems to persist in size and shape as the number of spins
increases, it is currently unclear whether it survives in the
thermodynamic limit. In addition a region (labeled `4P' in Figure
\ref{fig:J4}(b)) where the ground state is unpolarized but different
from the antiferromagnetic and dimer phases occurs. This `4P' region
could correspond to a single or several phases. Unfortunately, the
size dependence in this part of the phase diagram turns out to be very
complicated. Due to frustration introduced by the four-particle
exchange, a large number of low-energy states exist. Therefore, the
study of short chains does not allow one to determine the properties
of the ground state in this regime.

\subsection{Spin phases of interacting quantum wires in the
  quasi-one-dimensional regime}
\label{sec:prefactors}

While the above results were obtained in the limit $r_\Omega\gg1$,
interaction parameters in realistic quantum wires vary widely, ranging
from $r_\Omega<1$ in cleaved-edge overgrowth wires to $r_\Omega\approx
20$ in p-type gate-defined wires~\cite{Danneau,klochan,Danneau2}.
While the former are weakly interacting, the latter are clearly in the
strongly interacting regime. However, the above analysis based solely
on exponents is not sufficient to determine the ground state of
interacting electrons in a quantum wire at finite $r_\Omega$. In order
to obtain a phase diagram in that case, the prefactors $J_l^*$ have to
be computed which can be done by including Gaussian fluctuations
around the classical exchange paths.

\begin{figure}[tb]
  \hspace*{0.16\textwidth}
  \resizebox{0.5\textwidth}{!}{\includegraphics{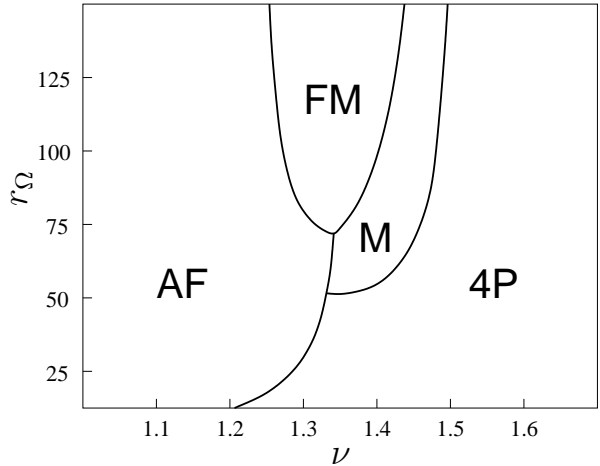}}
  \caption{\label{fig:phases-QW} Spin ground states of interacting
    electrons in quantum wires in the zigzag regime~\cite{kmhm}.}
\end{figure}

Using the exchange constants $J_l(\nu,r_\Omega)$ computed in this
way~\cite{kmhm} and the phase diagram shown in Figure \ref{fig:J4},
the ground states realized for given system parameters can be
determined. The resulting phase diagram is shown in Figure
\ref{fig:phases-QW}. It turns out that the partially and fully
polarized phases are realized only at large $r_\Omega\gtrsim 50$. At
moderately large $r_\Omega$ the transition occurs directly from the
antiferromagnetic phase to a phase dominated by the four-particle ring
exchange. These findings, thus, do not support the interpretation
\cite{Thomas, Thomas_1, Thomas_2, Crook, Kristensen, Kane, Reilly_1,
  Rokhinson, Berggren1, Berggren2, Berggren3, Spivak, Reilly} of the
so-called 0.7 anomaly in terms of spontaneous spin polarization.

Even at sufficiently strong interactions, the question arises of how a
ferromagnetic state in the quantum wire manifests itself in the
conductance.  It is tempting to assume that in the fully polarized
state the wire supports only one excitation mode and thus has
conductance $e^2/h$.  This is indeed the case when the full
polarization of electron spins is achieved by applying a sufficiently
strong magnetic field.  Such a field creates a gap in the spectrum of
spin excitations, and below the gap the system is equivalent to a
spinless electron liquid with conductance $e^2/h$.  It is important to
stress that the situation is very different if the full spin
polarization is achieved due to internal exchange processes in the
electron system, rather than the external magnetic field.  In this
case, the ground state is degenerate with respect to spin rotations,
and thus the system supports gapless spin excitations---the magnons.
As a result, the conventional argument in favor of conductance value
$e^2/h$ no longer applies.

In studying the conductance of a ferromagnetic wire it is important to
keep in mind that the properties of the electron system inside the
quantum wire in general do not fully determine its conductance.
Indeed, since the electric current flows between non-magnetic leads
through a ferromagnetic wire, the spatial non-uniformity of the system
needs to be considered carefully, and the problem of determining the
conductance complicates considerably.  In the case of a ferromagnetic
zigzag Wigner crystal in the middle of the wire, the weakening of the
confining potential in the contact region would lead to either melting
of the crystal or the emergence of a crystal with more and more rows.
In both cases modeling of the spin interactions in the transition
region is by no means obvious.

\begin{figure}[tb]
  \hspace*{0.16\textwidth}
  \resizebox{0.5\textwidth}{!}{\includegraphics{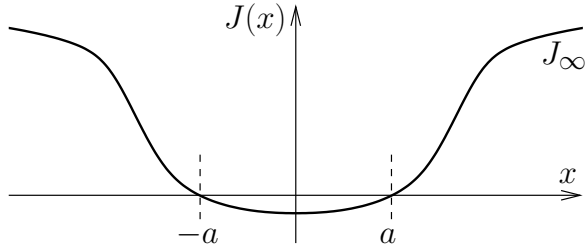}}
  \caption{\label{fig:cond-FM} Simple model of exchange coupling in a
    ferromagnetic Wigner crystal coupled to non-magnetic leads.  The
    coupling constant vanishes in the contact region at points $-a$
    and $a$, and therefore the spin excitations in the leads and in
    the wire decouple.}
\end{figure}

The simplest model that might capture the relevant physics is one
where the system is described by Hamiltonian (\ref{eq:spin-chain})
with an effective position-dependent nearest-neighbor exchange
constant $J(x)$ as depicted in Figure \ref{fig:cond-FM}. In the leads,
interactions are weak and antiferromagnetic and therefore $J$ is large
and positive. In the wire, interactions are strong and ferromagnetic
and therefore $J$ is small and negative. Through the contact regions,
$J$ varies smoothly and changes sign at points $-a$ and $a$.  Within
this model, the arguments of section \ref{sec:conductance} lead to the
conclusion that the spin polarization does suppress the conductance.
Namely, since the exchange coupling constant vanishes at the borders
of the ferromagnetic region, i.e., at $\pm a$, the spin degrees of
freedom in the leads are decoupled from those in the wire and, thus,
the propagation of spin excitations through the wire is blocked.
Accordingly, the value of the conductance is reduced by a factor $2$.
By contrast to the antiferromagnetic case in one-dimensional wires,
this suppression would persist down to temperatures $T\to0$ due to the
vanishing of the spin coupling in the contact region.

Of course the contact region in real quantum wires is more complex and
a satisfactory theory for the conductance of strongly interacting
quasi-one-dimensional quantum wires that correctly takes into account
the spin degrees of freedom is an open problem.

\section{Orbital properties of zigzag Wigner crystals}
\label{sec:twomode}

In addition to the spin physics discussed in the previous section,
quasi-one-dimensional wires have interesting orbital properties.  In
this section, we discuss the transition from a one-dimensional to a
quasi-one-dimensional state for the case of spinless electrons, based
mainly on Ref.~\cite{MML}.

As shown in section \ref{sec:classical_zigzag}, the classical
transition from a one-dimensional to a zigzag Wigner crystal can be
obtained simply by minimizing the energy of the interacting electron
system in a transverse confining potential.  A different way of
studying the same transition is by considering the phonon modes of the
crystal. In the one-dimensional crystal one longitudinal and one
transverse phonon mode exist.  The longitudinal phonon is gapless
because sliding of the entire crystal along the wire axis does not
cost any energy.  On the other hand, the transverse mode is gapped
with a gap frequency equal to the frequency of the confining
potential. The transition to a zigzag state is driven by a softening
of the transverse phonon at wave vector $k=\pi n$ which corresponds to
a staggered displacement of electrons transverse to the wire axis. In
the zigzag crystal, we obtain two longitudinal and two transverse
phonon modes with the following low-$q$ dispersions~\cite{MK-unpub}
close to the transition ($\delta\nu/\nu\ll1$):
\begin{eqnarray}
  \label{eq:phonon-longitudinal}
  \omega_{\parallel
    0}(q)=\frac\pi2\,\Omega\sqrt{\nu_c^3\ln\frac1{|q|}}\;|q|, \qquad
  \omega_{\parallel\pi}(q)=\sqrt2\,\Omega+{\cal 
    O}(q^2),\\
  \label{eq:phonon-transversal}
  \omega_{\perp 0}(q)=\Omega+{\cal O}(q^2),\qquad\omega_{\perp\pi}(q)
  =\sqrt 6\,\Omega\sqrt{\frac{\delta\nu}{\nu_c}+\frac{\pi^2\ln2}{48}\nu_c^3\,q^2},
\end{eqnarray} 
where $q=k/(\pi n)$.

Thus, only at the transition point $\delta\nu=0$, two gapless phonon
modes exist with dispersions $\omega_{\parallel0}(q)$ and
$\omega_{\perp\pi}(q)=(\pi/2\sqrt2)\,\Omega\sqrt{\nu_c^3\ln2}\,|q|$.
Within the zigzag regime, there is a single gapless excitation
corresponding to in-phase longitudinal motion of the two rows that
constitute the zigzag crystal. The soft-mode $\omega_{\perp\pi}(q)$
describing the out-of-phase transverse motion acquires a gap
$\Delta_{\rm cl}\propto\sqrt{\delta\nu}$.

This behavior is markedly different from the noninteracting case.  In
a noninteracting system the transition from a one-dimensional to a
quasi-one-dimensional state happens when the chemical potential is
raised above the subband energy of the second subband of transverse
quantization.  In the quasi-one-dimensional state, the two occupied
subbands are decoupled and each of them supports a gapless electronic
excitation mode, i.e., above the transition two gapless modes exist
rather than just one as in the classical Wigner crystal.

\begin{figure}[tb]
  \resizebox{0.32\textwidth}{!}{\includegraphics{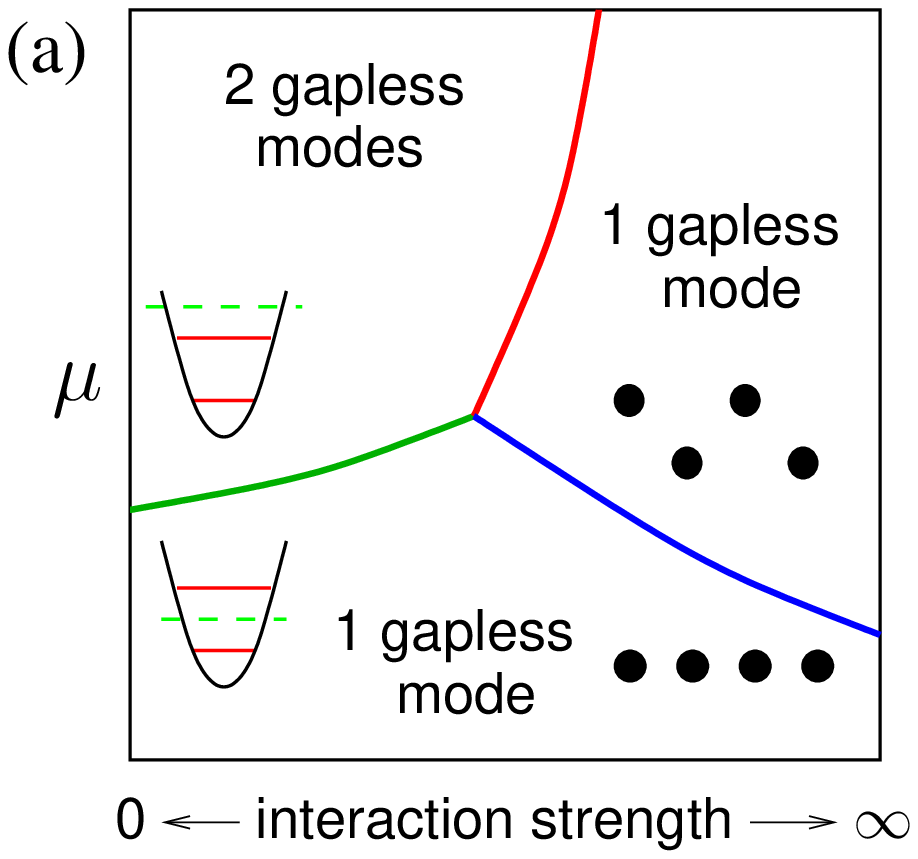}}
  \hfill
  \resizebox{0.32\textwidth}{!}{\includegraphics{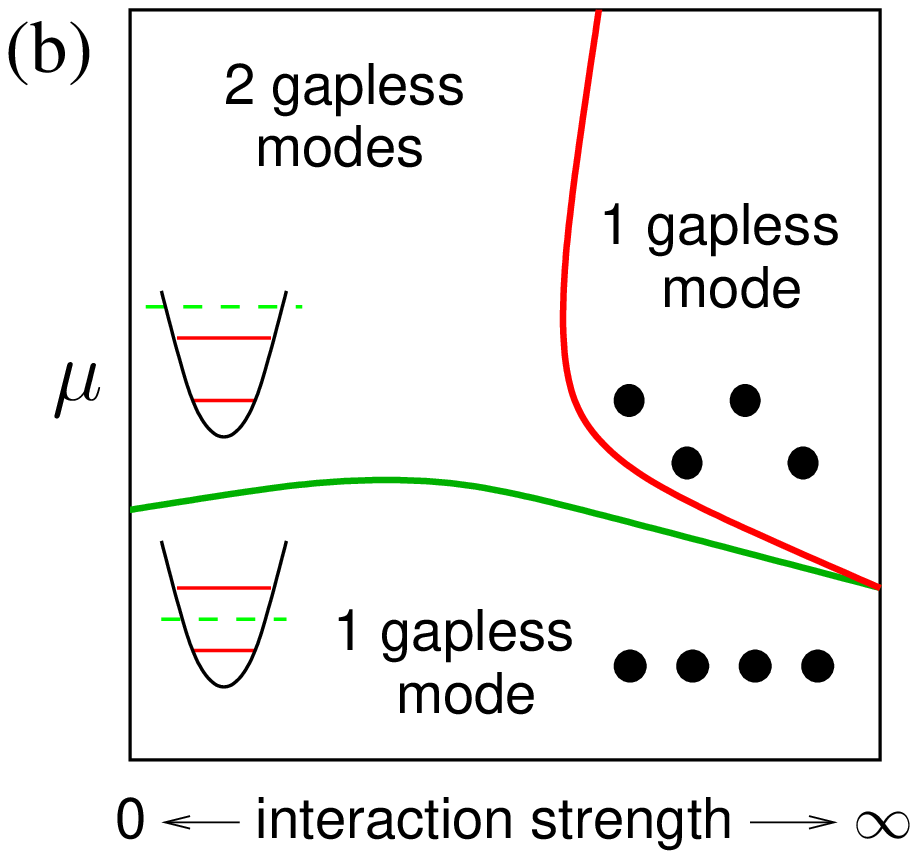}}
  \hfill
  \resizebox{0.32\textwidth}{!}{\includegraphics{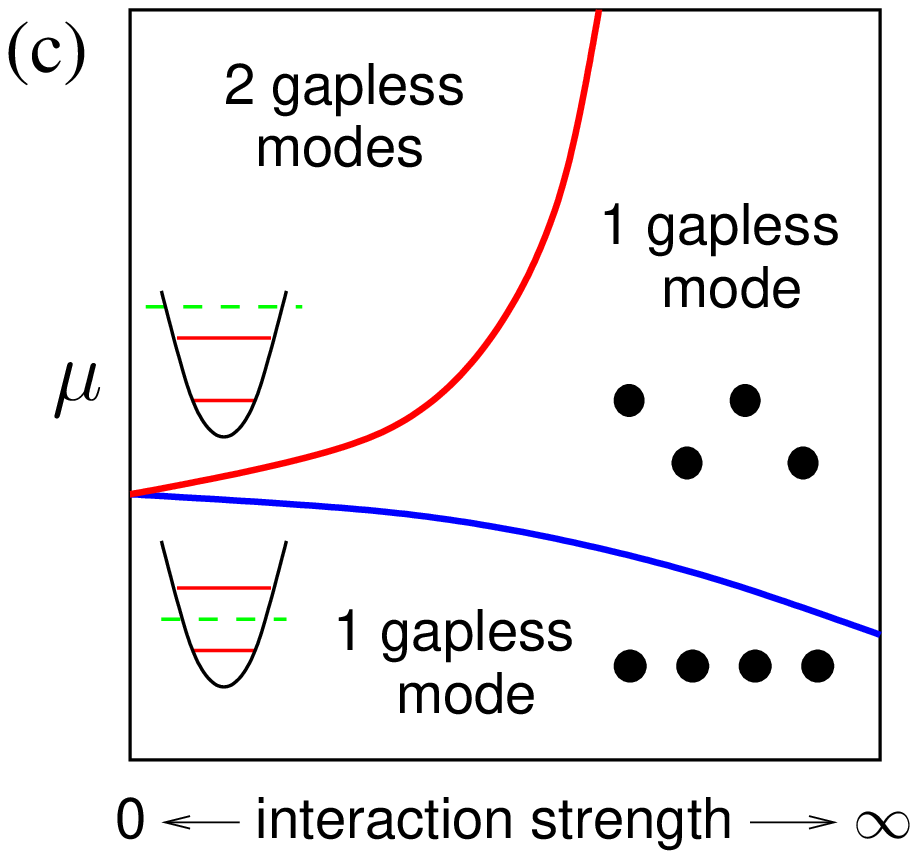}}
  \caption{\label{fig:pd1} At zero and infinite interaction strength
    the quasi-one-dimensional system supports two and one gapless
    excitation mode, respectively. Here the interaction strength is
    characterized by the parameter $r_\Omega$ introduced in
    Sec.~\ref{sec:classical_zigzag}.  Possible phase diagrams
    consistent with these findings are shown: (a) A tricritical point
    exists at a finite interaction strength. (b) Weak quantum
    fluctuations destroy the gap of the classical Wigner crystal.  (c)
    Already infinitesimally weak interactions induce a gap in the
    second mode.}
\end{figure}

One might, thus, expect that the phase diagram of the system as a
function of interaction strength is as shown in
Figure~\ref{fig:pd1}(a), namely two distinct quasi-one-dimensional
phases exist at weak and strong interactions. Consequently one should
find a tricritical point at a finite interaction strength where the
nature of the transition from a one-dimensional to a
quasi-one-dimensional state changes.  However, the phase diagram
Figure~\ref{fig:pd1}(a) is not the only one consistent with both the
above findings for the noninteracting case and the classical Wigner
crystal at infinite interaction strength. Two alternatives are shown
in Figure~\ref{fig:pd1}(b,c). To distinguish between the different
possibilities, one needs to study the nature of the transition as a
function of interaction strength. In particular, the following
questions have to be answered: {(i)} Do weak quantum fluctuations
destroy the gap found in the classical zigzag crystal at strong
interactions?  {(ii)} Do infinitesimally weak interactions lead to a
gap in the second mode above the transition?

\subsection{Quantum theory of the zigzag transition}
\label{sec:zigzag-quantum}

Let us consider the strongly interacting case first and account for
the quantum nature of the system. In particular, using the classical
Wigner crystal configuration as a starting point, we now include
quantum fluctuations.  The phonon modes (\ref{eq:phonon-longitudinal})
and (\ref{eq:phonon-transversal}) reflect the fact that there are only
two types of possible low-energy excitations: the longitudinal plasmon
mode and a staggered transverse mode.  It turns out that in the
vicinity of the transition the two modes decouple.  The acoustic
spectrum of the plasmon mode is protected by translational invariance
and, thus, at least one gapless excitation mode exists in the system.
More interesting is the staggered transverse mode.

To describe the transverse displacements of the electrons, a staggered
field $\varphi_l=(-1)^ly_l$ is introduced. In the vicinity of the
transition, $\varphi_l$ is slowly varying on the scale of the
inter-electron distances and, therefore, the continuum limit
$\varphi_l\to\varphi(x)$ can be taken. Expanding the action up to
fourth order in $\varphi$, one finds
\begin{eqnarray}
  \label{eq:phi4}
  S[\varphi]\eq A\hbar\sqrt{r_\Omega}\int d\tau\,dx\left[(\partial_\tau\varphi)^2+
    (\partial_x\varphi)^2-\delta\nu\,\varphi^2+\varphi^4\right],
\end{eqnarray}
where the variables have been rescaled such as to provide the simplest
action possible. The form of the action as well as all following
conclusions do not depend on the exact shape of the confining
potential.  For a parabolic confining potential the constant $A$ is
given as $A=[7\zeta(3)]^{3/2}\sqrt{\ln2}/(31\zeta(5))$.

\begin{figure}[tb]
  \hspace*{0.16\textwidth}
  \resizebox{0.75\textwidth}{!}{\includegraphics{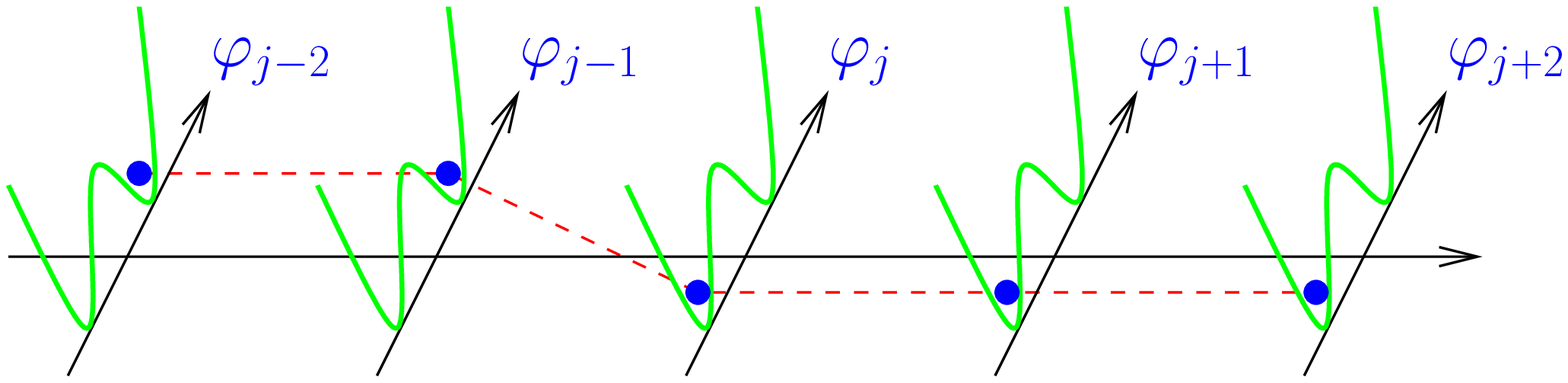}}
  \caption{\label{fig:spin-chain} Mapping of the $\varphi^4$-theory to
    a spin chain.}
\end{figure}

The classical transition point as discussed above corresponds to
$\delta\nu=0$.  Here the transverse mode becomes unstable, and the
quartic term is needed to stabilize the system. Quantum fluctuations
may affect both the transition point and the nature of the transition.
A convenient way to analyze the quantum-mechanical problem is to
refermionize. As a first step, we rediscretize the coordinate $x$
along the wire axis.  The discrete version of the Hamiltonian then
describes a set of particles moving in a double-well potential $V_{\rm
  DW}\sim-\lambda\varphi_j^2+\varphi_j^4$ (as depicted in Figure
\ref{fig:spin-chain}) and interacting through a nearest-neighbor
interaction $\propto (\varphi_j-\varphi_{j+1})^2$.  If the double-well
potential is sufficiently deep, i.e., if $\lambda$ is sufficiently
large, the particles are almost completely localized in one of the
wells at $\varphi_j=\pm\sqrt{\lambda/2}$. Then each particle can be
described by a pseudo-spin operator, namely
$\varphi_j=\sqrt{\lambda/2}\,\sigma_j^z$, where $\sigma_j^z$ is a
Pauli matrix. In terms of these new variables, the Hamiltonian
consists of two terms corresponding to tunneling between the two wells
and the nearest-neighbor interaction, respectively. Tunneling is
described by $H_t=-t\sum_j\sigma_j^x$ whereas the interaction term
reads $H_{\rm NN}=-v\sum_j\sigma_j^z\sigma_{j+1}^z$. The resulting
Hamiltonian $H_t+H_{\rm NN}$ is the Hamiltonian of the transverse
field Ising model~\cite{Tsvelik}. Here the parameters $t$ and $v$ are
related to the parameters of the original model. In particular, they
can be tuned by changing the chemical potential which controls the
transition, i.e., $t=t(\mu)$ and $v=v(\mu)$.

In order to arrive at a fermionic description, a Jordan-Wigner
transformation is used. It turns out that the Hamiltonian takes a much
simpler form if one rotates $\sigma^x\to-\sigma^z$ first. The
representation of the spin operators in terms of (spinless) fermions,
\begin{equation}
  \fl
  \sigma_j^+\equiv \sigma_j^x+i\sigma_j^y
  =2a_j^\dagger \rme^{\rmi\pi\sum_{i<j}a_i^\dagger
    a_i},
  \quad
  \sigma_j^-\equiv \sigma_j^x-i\sigma_j^y=2\rme^{-\rmi\pi\sum_{i<j}a_i^\dagger
    a_i}a_j,\quad\sigma_j^z=2a_j^\dagger a_j-1,
\end{equation}
where $a^\dagger_j$ and $a_j$ are fermion creation and annihilation
operators on site $j$, respectively, then yields the noninteracting
Hamiltonian~\cite{Tsvelik}
\begin{eqnarray}
  H_{\rm f}=\sum_j\Big[2t\,a_j^\dagger
  a_j-v\Big(a_j^\dagger-a_j\Big)\Big(a_{j+1}^\dagger+a_{j+1}\Big)\Big].
  \label{eq-JW}
\end{eqnarray}
Note that this Hamiltonian is essentially the transfer matrix of the
two-dimensional classical Ising model~\cite{Mattis} near the
transition.  The connection can be made clear by considering the
mapping between $d$-dimensional quantum and $(d+1)$-dimensional
classical models~\cite{Vaks-Larkin}. Thus, the one-dimensional quantum
Ising model studied here is equivalent to the two-dimensional
classical Ising model~\cite{Polyakov}.

In Hamiltonian (\ref{eq-JW}) one can identify three different
contributions: $-v(a_j^\dagger a_{j+1}+a_{j+1}^\dagger a_j)$ describes
a tight-binding model with bandwidth $4v$, the local term
$2ta_j^\dagger a_j$ yields the chemical potential $\mu_f=-2t$ of the
spinless fermions, and finally $-v(a_j^\dagger
a_{j+1}^\dagger-a_ja_{j+1})$ is a BCS-like pairing term with $p$-wave
symmetry.  The one-dimensional regime of our original model
corresponds to $t>v$ when the chemical potential lies below the bottom
of the tight-binding band and, therefore, all the fermionic states are
empty.  The transition to a quasi-one-dimensional regime happens when
the chemical potential reaches the bottom of the band at $t=v$. For
$t<v$, some of the fermionic states describing the motion of the
original electrons transverse to the wire axis are filled. Due to the
pairing term in the Hamiltonian, these states acquire a gap.  With the
help of a Bogoliubov transformation, the Hamiltonian can be
diagonalized to obtain the energy spectrum. As a result one finds that
the gap is given as $\Delta=2|t-v|$.

\begin{figure}[tb]
  \hspace*{0.16\textwidth}
  \resizebox{0.5\textwidth}{!}{\includegraphics{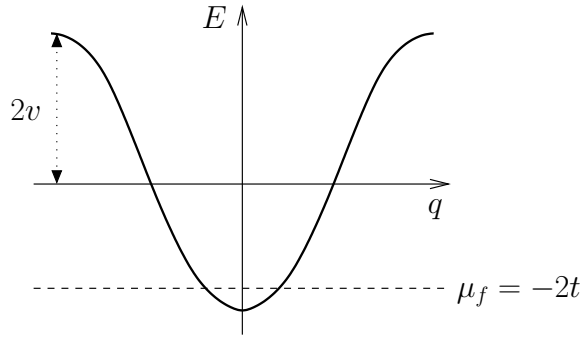}}
  \caption{\label{fig:tightbinding} After the Jordan-Wigner
    transformation, one obtains a tight binding model for spinless
    fermions with bandwidth $2v$ and chemical potential $-2t$. The
    transition from a one-dimensional to a quasi-one-dimensional state
    happens when the chemical potential reaches the bottom of the
    band.}
\end{figure}

In terms of the original system parameters, one expects that $t$ and
$v$ are non-singular functions of the chemical potential. The critical
chemical potential $\mu_c$ is thus defined by the condition
$t(\mu_c)=v(\mu_c)$. Furthermore, the behavior of the gap is obtained
by expanding $t-v$ in the vicinity of the transition point $\mu_c$. As
a consequence one obtains a linear gap,
\begin{equation}
  \label{eq:Ising_gap}
  \Delta\propto |\mu - \mu_c|.
\end{equation}
While quantum effects modify the nature of the transition, the number
of gapless excitations remains the same: in the strongly interacting
system, only one gapless excitation exists.  Thus, the phase diagram
of Figure~\ref{fig:pd1}(b) is ruled out---weak quantum fluctuations do
not destroy the gap of the transverse mode in the
quasi-one-dimensional regime. To differentiate between the phase
diagrams of Figures~\ref{fig:pd1}(a) and \ref{fig:pd1}(c), a
complementary approach has to be used.  In the next section we
consider the limit of weak interactions to check whether or not they
lead to the formation of a gap just above the transition, as suggested
by the scenario of Figure~\ref{fig:pd1}(c).

\subsection{Two-subband system at weak interactions}

The limit of weak interactions can be treated using a renormalization
group approach. Here the description in terms of two subbands due to
transverse size quantization is a good starting point. Each subband is
described by fermionic operators $\psi_j$, where $j=1,2$. The free
Hamiltonian is just
\begin{eqnarray}
  H_0\eq\sum_j\int dx\left[
    -\frac{\hbar^2}{2m}\psi^\dagger_j\partial^2\psi_j
    +\varepsilon_j\psi^\dagger_j\psi_j\right],
\end{eqnarray}
where $\varepsilon_j$ are the subband energies. For a parabolic
confining potential, $\varepsilon_j=\hbar\Omega(j-\frac12)$.

\begin{figure}[tb]
  \hfill \resizebox{\textwidth}{!}{\includegraphics{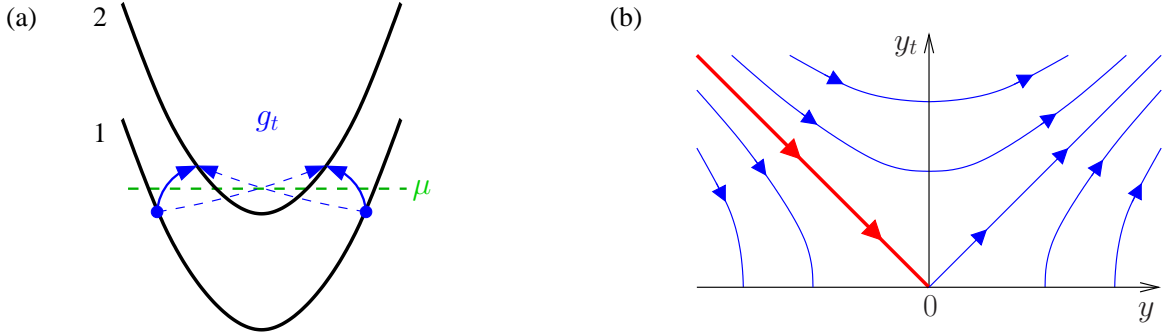}}
  \caption{\label{fig:gt} (a) Two-particle transfer between subbands
    described by the coupling constant $g_t$. (b) Weak coupling RG
    flow.}
\end{figure}

Interactions can be separated into intra-subband and inter-subband
interactions. One needs four interaction constants to describe the
system: $g_j\sim V_{jj}(0)-V_{jj}(2k_{Fj})$ which describes
intra-subband forward scattering,
\begin{eqnarray}
  g_x\sim V_{12}(0)-\frac12(V^{\rm
    ex}_{12}(k_{F1}-k_{F2})+V^{\rm ex}_{12}(k_{F1}+k_{F2}))
\end{eqnarray}
which describes inter-subband forward scattering, and
\begin{eqnarray}
  g_t\sim V^{\rm
    ex}_{12}(k_{F1}-k_{F2})-V^{\rm ex}_{12}(k_{F1}+k_{F2})
  \label{eq:gt0}
\end{eqnarray}
which describes transfer of two particles between the subbands as
shown in Figure \ref{fig:gt}(a). Here
\begin{eqnarray}
  V_{ij}(k)\eq\int dx\,dx'\;\rme^{\rmi k(x\!-\!x')}\int dy\,dy'\;V_{\rm
    int}({\bf r}\!-\!{\bf r}')\chi_i^2(y)\chi_j^2(y'),\\
  V_{ij}^{\rm
    ex}(k)\eq\int dx\,dx'\;\rme^{\rmi k(x\!-\!x')}\int dy\,dy'\;V_{\rm
    int}({\bf r}\!-\!{\bf r}')\chi_i(y)\chi_j(y)\chi_i(y')\chi_j(y'),
\end{eqnarray}
where $\chi_i$ are the transverse eigenmodes in the confining
potential.

It is well known that forward scattering in one dimension does not
open a gap in the system~\cite{giamarchi}. By contrast, the
two-particle transfer between subbands described by the coupling
constant $g_t$ could open a gap. To assess whether this is indeed the
case, the renormalization group (RG) is used, i.e., reducing the
bandwidth from $D_0$ down to $D$, the scale-dependent coupling
constants are determined. The RG equations for a two-band system are
given as~\cite{RG1,RG2,Ledermann}
\begin{eqnarray}
  g_1'\eq-\frac1{2\pi\hbar v_{F2}}g_t^2,\qquad g_2'= -\frac1{2\pi\hbar
    v_{F1}}g_t^2,\qquad 
  g_x'=\frac1{\pi\hbar(v_{F1}\!+\!v_{F2})}g_t^2,\\
  g_t'\eq-\frac1{2\pi\hbar}\left(\frac{g_1}{v_{F1}}+\frac{g_2}{v_{F2}}
    -\frac{4g_x}{v_{F1}\!+\!v_{F2}}\right)g_t,
\end{eqnarray}
where the derivatives are taken with respect to $\xi=\ln(D_0/D)$.  By
introducing the dimensionless coupling constants
\begin{eqnarray}
  \fl y=-\frac1{2\pi\hbar } \left(\frac{g_1}{v_{F1}} + \frac{g_2}{v_{F2}} -
    \frac{4g_x}{v_{F1}+v_{F2}}\right),\qquad y_t= \frac{g_t}{\pi\hbar} 
  \sqrt{\frac{(v_{F1}+v_{F2})^2+4v_{F1}v_{F2}}
    {2v_{F1}v_{F2}(v_{F1}+v_{F2})^2}},
\end{eqnarray}
the four RG equations can be combined into two
equations~\cite{Ledermann}
\begin{eqnarray}
  y'= y_t^2,\qquad\qquad y_t'= yy_t.\end{eqnarray}
The flow diagram corresponding to these equations is shown in Figure
\ref{fig:gt}(b). 

The coupling constant $g_t$ describes a combination of two processes:
the particles transferred between subbands may retain their direction
of motion, corresponding to momentum transfer $\pm(k_{F1}-k_{F2})$, or
they may change their direction of motion, corresponding to momentum
transfer $\pm(k_{F1}+k_{F2})$. Both processes are depicted in Figure
\ref{fig:gt}(a). The resulting coupling constant, thus, is
proportional to $V^{\rm ex}_{12}(k_{F1}-k_{F2})-V^{\rm
  ex}_{12}(k_{F1}+k_{F2})$ as given in equation~(\ref{eq:gt0}).
Consequently, $g_t^{(0)}\propto k_{F2}$ as the density in the second
subband goes to zero, and one concludes that $y_t^{(0)}\propto
\sqrt{v_{F2}}$ is much smaller that $y^{(0)}$.  Therefore the presence
or absence of a gap just above the transition to a
quasi-one-dimensional state is determined by the sign of $y^{(0)}$.
If $y^{(0)}<0$, the coupling constant $y_t$ flows to zero, and the
system remains gapless. On the other hand, if $y^{(0)}>0$, the
coupling constant $y_t$ flows to infinity, and the system acquires a
gap.

The interaction constant $y^{(0)}$ can be evaluated assuming a Coulomb
interaction screened by a gate at a distance $d$ much larger than the
effective width of the wire, i.e.,
\begin{eqnarray}
  V_{\rm int}(r) \eq \frac{e^2}{2\epsilon}
  \left(\frac1{r}-\frac1{\sqrt{r^2+(2d)^2}}\right) 
\end{eqnarray}
which is the two-dimensional version of
Eq.~(\ref{eq:screened_interaction}).
 
One finds that $g_1\approx g_x\approx2 (e^2/\epsilon)\ln k_{F1}d$
(with logarithmic accuracy). On the other hand, at densities
$n_2\ll1/d$,
\begin{eqnarray}
  g_2\sim\frac{e^2}\epsilon(k_{F2}d)^2\ln\frac1{k_{F2}d},
\end{eqnarray}
i.e., the interaction constant $g_2$ vanishes in the limit $n_2\to0$.
This is a manifestation of the Pauli exclusion principle: Once the
average distance between particles exceeds the distance to the gate,
the interactions become effectively local. However, identical fermions
do not interact via a local interaction, hence $g_2\to0$.

Using the above expressions for the interaction constants, one finds
the initial value $y^{(0)}\approx3g_1/(2\pi\hbar v_{F1})>0$. As
$y^{(0)}$ is positive, the system flows to strong coupling.  Thus, the
system develops a gap close to the transition from a one-dimensional
to a quasi-one-dimensional state, and therefore the phase diagram
Figure~\ref{fig:pd1}(a) does not describe the system.

\begin{figure}[tb]
  \hspace*{0.16\textwidth}
  \resizebox{0.45\textwidth}{!}{\includegraphics{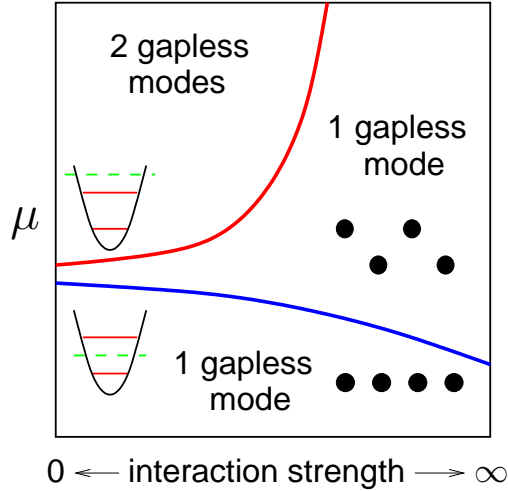}}
  \caption{\label{fig:variational} Phase diagram of spinless
    interacting electrons in a quantum wire~\cite{MML}. At any
    interaction strength there is a finite window of densities where
    the quasi-one-dimensional system supports only one gapless
    excitation.}
\end{figure}

A second gapless excitation mode appears only once the density is
increased further beyond the transition point. As the density
increases, $g_2/v_{F2}$ increases and becomes comparable to
$g_1/v_{F1}$. Then $y^{(0)}$ changes sign and eventually one crosses
into the regime where $y_t$ scales to zero. At weak interactions this
happens at $k_{F2}\sim 1/(k_{F1}d^2)$.  Thus, at any interaction
strength there is a finite window of densities in which the system is
in the quasi-one-dimensional state but supports only one gapless
excitation mode.  The resulting phase diagram~\cite{MML} is shown in
Figure \ref{fig:variational}.

Having found that the behavior at weak and strong interactions is very
similar, there is no reason to expect that at intermediate
interactions no gap exists. One notices, however, that the magnitude
of the gap strongly depends on interaction strength. In the Wigner
crystal we find a large Ising gap $\Delta\sim|\delta\mu|$. At weak
interactions, on the other hand, the gap scales with a large exponent,
$\Delta\sim(\delta\mu)^{\alpha}$, where $\alpha=1/(4y^{(0)})$.

\subsection{Intermediate interactions}

The method of choice to treat intermediate interactions in
one-dimensional and quasi-one-dimensional systems is bosonization.
However, bosonization requires a linear spectrum. In the present case
this is not straightforward because, to describe the transition, one
is necessarily interested in what happens at the bottom of the second
subband where a linearization is not justified.

Alternatively one may bosonize the first subband which has a large
Fermi energy and keep a fermionic description in the second
subband~\cite{balents}. Thus, the electrons in the first subband are
described by the bosonic fields $\phi_1(x)$ and $\theta_1(x)$ whereas
the electrons in the second subband are described by the fermionic
creation and annihilation operators, $\psi_2^\dagger(x)$ and
$\psi_2(x)$. In particular,
\begin{eqnarray}
  \fl H\eq\frac{\hbar
    v_{F1}}{2\pi}\int dx\,\left((\partial\theta_1)^2+
    \frac1{K^2}(\partial\phi_1)^2\right)-\frac{\hbar^2}{2m}\int
  dx\;\psi_2^\dagger\partial^2\psi_2\nn\\
  \fl && -\frac1\pi\sum_q (\partial\phi_1)(q) V_{12}(q)n_2(-q) + \gamma_t\int
  dx\,\Big[\big(\psi_2^\dagger\partial\psi_2^\dagger
  - \partial\psi_2^\dagger\psi_2^\dagger\big) \rme^{-2\rmi\theta_1} + {\rm h.c.}\Big],\label{eq:before-UT}
\end{eqnarray}
where $\gamma_t\sim e^2/\epsilon$.  Furthermore, $K=(1+g_1/\pi\hbar
v_{F1})^{-1/2}$ is the Luttinger parameter in the first subband.

The dominant interaction between the bosons and the fermions is the
inter-subband forward scattering $V_{12}$. This coupling can be
eliminated by applying a unitary transformation
\begin{eqnarray}
  U=\exp\left[-\frac {\rmi K^2}{\pi\hbar v_{F1}} \int
    dx\,dy\;\theta_1(x)V(x-y)n_2(y)\right].
  \label{eq:UT}
\end{eqnarray}
The new Hamiltonian in terms of the transformed bosonic
($\phi_c,\theta_c$) and fermionic ($\psi_s^\dagger,\psi_s$) fields
then reads
\begin{eqnarray}
  H_U=UHU^\dagger\eq\frac{\hbar
    v_{F1}}{2\pi}\int dx\,\left((\partial\theta_c)^2 +
    \frac1{K^2}(\partial\phi_c)^2\right)-\frac{\hbar^2}{2m}\int
  dx\;\psi^\dagger_s
  \partial^2\psi_s\nn\\
  &&+\gamma_t\int
  dx\,\Big[\big(\psi_s^\dagger\partial\psi_s^\dagger
  - \partial\psi_s^\dagger\psi_s^\dagger\big)\rme^{-2\rmi\kappa\theta_c(x)} + {\rm h.c.}\Big],\label{eq:after-UT}
\end{eqnarray}
i.e., the inter-subband forward scattering disappears. Comparing the
Hamiltonians (\ref{eq:before-UT}) and (\ref{eq:after-UT}), note that
the exponent in the boson-fermion interaction term changes from
$2\rmi\theta_c$ to $2\rmi\kappa\theta_c$, where
$\kappa=1-K^2g_x/(\pi\hbar v_{F1})\simeq K^2$. It is essential to
realize that other than that the Hamiltonian preserves its form after
the unitary transformation.  Namely we are still dealing with a
plasmon mode coupled to noninteracting fermions. Not only the bare
interaction in the second subband, but also the effective interaction
generated by the inter-subband forward scattering vanishes in the
limit $n_2\to0$.  Additional terms that are generated by the unitary
transformation can be shown to be irrelevant \cite{balents}.

Since the fermions remain noninteracting, as a next step, they can be
bosonized. The purely bosonic Hamiltonian could then in principle be
subject to an RG approach. Or, as it is safe to assume that at
intermediate interactions the second mode is still gapped, one may use
a variational approach instead. One finds that the gap exponent
decreases with increasing interaction strength until the variational
approach is no longer valid because the relevant energy scale exceeds
the Fermi energy in the second subband~\cite{MK-unpub}. For stronger
interactions, note that the Hamiltonian (\ref{eq:after-UT}) in the
limit $K\to0$ takes the same form as the Hamiltonian of the Wigner
crystal with a gapless plasmon mode decoupled from the gapped Ising
fermions.\footnote{A more careful treatment~\cite{unpub} shows that at
  strong interactions the weak coupling of the two modes is marginally
  irrelevant and leads to relatively insignificant corrections to the
  Ising picture of the transition.}

\subsection{Experimentally observable consequences}

As mentioned earlier the computation of observables such as the
conductance is complicated due to the importance of the coupling to
leads.  One may speculate, however, how the above findings affect
observables.

The experimentally most relevant observable is the conductance. For
noninteracting electrons, the second subband opens a new channel in
the wire and, therefore, at the transition the conductance doubles
from $G=e^2/h$ to $G=2e^2/h$ (for spinless electrons). In the
interacting case, however, the second mode is gapped. One might
argue~\cite{starykh} that the total charge mode (plasmon) remains
gapless and, therefore, one should still expect a doubling of the
conductance at the transition. However, as discussed in section
\ref{sec:1D_crystal}, the conductance is not determined by the total
charge mode only. As the wire is coupled to leads, mixing between
different channels occurs and, therefore, modes other than the total
charge mode do affect the conductance. We expect that in this case,
too, the fact that the second mode is gapped leads to a suppression of
the conductance which should remain at its one-dimensional value of
$G=e^2/h$ until the second mode becomes gapless at a higher density.
This means that the transition from a one-dimensional state to a
quasi-one-dimensional state and the step in conductance no longer
coincide.  Only at higher temperatures $T>\Delta$ does the gapped mode
open for transport. Thus, the presence of a gapped mode is expected to
lead to non-trivial temperature dependence of the conductance.

\begin{figure}[tb]
  \hfill \resizebox{\textwidth}{!}{\includegraphics{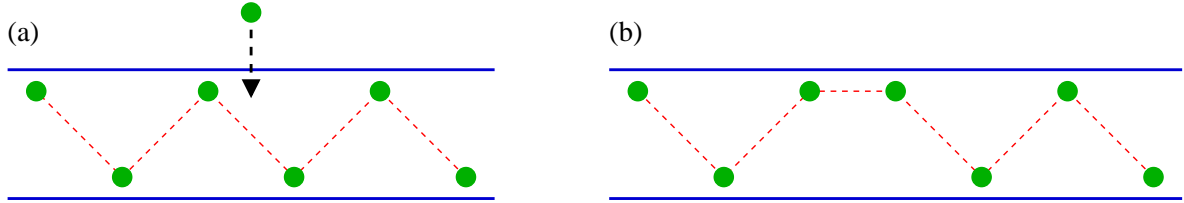}}
  \caption{\label{fig:DOS} (a) Tunneling into a zigzag crystal.  (b)
    In addition to exciting the plasmon mode, the tunneling electron
    creates a defect. The finite energy cost associated with the
    process manifests itself in a gap in the tunneling density of
    states.}
\end{figure}

The gapped mode above the transition should manifest itself most
clearly in the tunneling density of states.  Consider the Wigner
crystal limit. In the one-dimensional case, the addition of an
electron to the system requires the excitation of the plasmon mode in
order to adjust the density along the wire. Due to the stiffness of
the plasmon mode, the tunneling density of states is suppressed: As
discussed in section \ref{sec:structure} the Wigner crystal described
as an elastic medium can be viewed as a Luttinger liquid, and the
tunneling density of states of a Luttinger liquid is well known to
display a power-law suppression at the Fermi level
\cite{LL-DOS-1,LL-DOS-2}.  In the zigzag crystal, the addition of an
electron to the system also requires to adjust the density along the
wire by exciting plasmons which suggests a power-law suppression of
the density of states.  Apart from that, however, the addition of an
electron creates a defect in the zigzag structure: the electron is
added to one of the two rows and, thus, interrupts the zigzag pattern
as depicted in Figure \ref{fig:DOS}. The energy of such a defect is
finite, and, therefore, the density of states acquires a gap.

This behavior is not limited to the Wigner crystal. As discussed in
Ref.~\cite{starykh}, at any interaction strength tunneling of a single
electron into the bulk of the wire excites both the gapless and the
gapped mode,\footnote{Within the formulation presented here, this can
  be understood by going back to the unitary transformation
  (\ref{eq:UT}) which relates the original degrees of freedom
  described by Hamiltonian (\ref{eq:before-UT}) to the new degrees of
  freedom described by Hamiltonian (\ref{eq:after-UT}). In particular,
  applying the unitary transformation to the single electron creation
  operators $\psi_j^\dagger$, where $j=1,2$ is the subband index, one
  may verify that $\psi_1^\dagger$ as well as $\psi_2^\dagger$ contain
  contributions from both the gapless and the gapped mode of
  (\ref{eq:after-UT}).  } and the finite energy cost associated with
excitation of the gapped mode entails a gap in the tunneling density
of states. Consequently the observation of a gap opening in the
tunneling density of states would allow one to identify the transition
to the quasi-one-dimensional state.

\section{Conclusion}
\label{sec:conclusion}

The Luttinger liquid physics of one-dimensional electron systems with
weak to moderate interactions has been studied extensively. The
present review focuses on novel phenomena due to \emph{strong}
interactions which lead to the formation of a Wigner
crystal~\cite{Wigner}. The strongly interacting regime can be realized
experimentally, and evidence for Wigner crystal physics has been seen
in the conductance of quantum wires~\cite{exp1,exp2}, the Coulomb
blockade peaks in carbon nanotubes~\cite{CNT}, and
possibly~\cite{qmc,baranger-0807.4292} in the localization features in
double quantum wires~\cite{yacoby-double-wires1,yacoby-double-wires2}.
While no phase transition takes place, at stronger interactions the
system properties change due to the presence of two very different
energy scales, namely the Fermi energy $E_F$ and the spin exchange
energy $J\ll E_F$. In the strictly one-dimensional regime, one of the
main features of Luttinger liquid physics is spin-charge
separation~\cite{giamarchi}. In an inhomogeneous Wigner-crystal wire,
however, spin physics is found to affect the conductance, reducing it
from $2e^2/h$ at $T\ll J$ to $e^2/h$ in the temperature regime $J\ll
T\ll E_F$ as discussed in Sec.~\ref{sec:1D_crystal}.

Real systems are never strictly one-dimensional, but confined by an
external potential. The presence of transverse degrees of freedom
leads to a transition from a one-dimensional to a zigzag Wigner
crystal at a finite electron density, see
Sec.~\ref{sec:classical_zigzag}. In contrast to the one-dimensional
crystal, the zigzag Wigner crystal displays a variety of spin ground
states as a function of density, see Sec.~\ref{sec:spins}.  In
particular, a ferromagnetic ground state---which has been suggested as
a possible cause of the conductance anomalies observed in quantum
wires \cite{Thomas}---can be realized.  While for noninteracting
electrons the transition to a quasi-one-dimensional state entails the
emergence of a second gapless excitation mode, this is not the case in
the presence of interactions.  As discussed in Sec.~\ref{sec:twomode},
the orbital degrees of freedom are strongly affected by interactions
which, for example, are expected to lead to a gap in the tunneling
density of states.  The most interesting properties of the
quasi-one-dimensional state in quantum wires are summarized in the
phase diagrams Figure \ref{fig:phases-QW} (spin properties) and Figure
\ref{fig:variational} (orbital properties).

Figure~\ref{fig:phases-QW} shows the spin phases of the zigzag Wigner
crystal obtained under the assumption that spin and orbital properties
can be treated separately.  This approach is justified at strong
interactions, when the energy scales for spin and charge excitations
are very different, $J\ll E_F$.  However, as interactions become
weaker the crystal starts to melt, leading to the coupling of spin and
orbital degrees of freedom.  To study the behavior of the system in
this regime, one needs to develop a theory that treats spin and
orbital degrees of freedom on equal footing.  This entails a number of
open questions: Are there remnants of the Wigner crystal phase
dominated by the four-particle ring exchange in the weakly interacting
quasi-one-dimensional state?  Is the spectral gap discussed in
Sec.~\ref{sec:twomode} robust to the inclusion of spin?

Figure~\ref{fig:variational} summarizes the orbital properties of the
electron system in a quantum wire near the transition from a
one-dimensional to a quasi-one-dimensional state.  The upper line
indicates the vanishing of the gap in the second excitation mode.
While the appearance of a second gapless excitation at a finite
distance above the transition has been shown in the limit of weak
interactions~\cite{MML}, more careful treatment is required to explore
this phenomenon at finite interaction strength.  In the opposite limit
of strong interactions, with increasing electron density the zigzag
regime eventually breaks down, giving way to structures with more than
two rows.  Numerical study of the quasi-one-dimensional Wigner
crystal~\cite{Piacente} shows that the number of rows changes as
$2\to4\to3\to4\to5$ \dots\ as a function of density.  Since in a wide
channel the electron density is not uniform across the
wire~\cite{Larkin-Shikin, Chklovskii}, this trend cannot persist up to
an arbitrarily high number of rows. Instead one expects that
structures with defects will have a lower energy.  The presence of
such potentially mobile defects will be crucial for understanding
transport properties of quantum wires in that regime.

\ack

This work was supported by the U. S. Department of Energy, Office of
Science, under Contract Nos.~DE-AC02-06CH11357 and DE-FG02-07ER46424.
We acknowledge our collaborators on various projects included in this
review: Akira Furusaki, Leonid Glazman, Toshiya Hikihara, Alexios
Klironomos, and Revaz Ramazashvili.  Furthermore, we thank the Aspen
Center for Physics, where part of this review was written, for
hospitality.


\section*{References}
\begin{thebibliography}{99}
  
\bibitem{Wees} van Wees B J, van Houten H, Beenakker C W J, Williamson
  J G, Kouwenhoven L P, van der Marel D and Foxon C T 1988
  \emph{Phys.\ Rev.\ Lett.}  \textbf{60} 848
  
\bibitem{Wharam} Wharam D A, Thornton T J, Newbury R, Pepper M, Ahmed
  H, Frost J E F, Hasko D G, Peacock D C, Ritchie D A and Jones G A C
  1988 \emph{J.\ Phys.} C \textbf{21} L209
  
\bibitem{Tarucha} Tarucha S, Honda T and Saku T 1995 \emph{Solid State
    Comm.}  {\bf 94} 413
  
\bibitem{Yacoby} Yacoby A, Stormer H L, Wingreen N S, Pfeiffer L N,
  Baldwin K W and West K W 1996 \emph{Phys. Rev. Lett.} {\bf 77} 4612
  
\bibitem{Tans} Tans S J, Devoret M H, Dai H, Thess A, Smalley R E,
  Geerligs L J, Dekker C 1997 \emph{Nature} \textbf{386} 474
  
\bibitem{Bockrath} Bockrath M, Cobden D H, McEuen P L, Chopra N G,
  Zettl A, Thess A, Smalley R E 1997 \emph{Science} \textbf{275} 1922
  
\bibitem{giamarchi} Giamarchi T 2004 \emph{Quantum Physics in One
    Dimension} (Oxford: Clarendon Press)
  
\bibitem{Landauer} Landauer R 1970 \emph{Phil. Mag.} \textbf{21} 863
  
\bibitem{maslov} Maslov D L and Stone M 1995 \emph{Phys. Rev.} B
  \textbf{52} R5539
  
\bibitem{ponomarenko} Ponomarenko V V 1995 \emph{Phys. Rev.} B
  \textbf{52} R8666
  
\bibitem{safi} Safi I and Schulz H J 1995 \emph{Phys. Rev.} B
  \textbf{52} R17040
  
\bibitem{Thomas} Thomas K J, Nicholls J T, Simmons M Y, Pepper M, Mace
  D R and Ritchie D A 1996 \emph{Phys.\ Rev.\ Lett.}  \textbf{77} 135
  
\bibitem{Thomas_1} Thomas K J, Nicholls J T, Appleyard N J, Simmons M
  Y, Pepper M, Mace D R, Tribe W R and Ritchie D A 1998 \emph{Phys.
    Rev.} B \textbf{58} 4846
  
\bibitem{Thomas_2} Thomas K J, Nicholls J T, Pepper M, Tribe W R,
  Simmons M Y and Ritchie D A 2000 \emph{Phys. Rev.} B \textbf{61}
  R13365
  
\bibitem{Crook} Crook R, Prance J, Thomas K J, Chorley S J, Farrer I,
  Ritchie D A, Pepper M and Smith C G 2006 \emph{Science} \textbf{312}
  1359
  
\bibitem{Kristensen} Kristensen A, Bruus H, Hansen A E, Jensen J B,
  Lindelof P E, Marckmann C J, Nyg\aa rd J, S\o rensen C B, Beuscher
  F, Forchel A and Michel M 2000 \emph{Phys. Rev.} B \textbf{62} 10950
  
\bibitem{Kane} Kane B E, Facer G R, Dzurak A S, Lumpkin N E, Clark R
  G, Pfeiffer L N and West K W 1998 \emph{Appl.\ Phys.\ Lett.}
  \textbf{72} 3506
  
\bibitem{Reilly_1} Reilly D J, Facer G R, Dzurak A S, Kane B E, Clark
  R G, Stiles P J, Clark R G, Hamilton A R, O'Brien J L, Lumpkin N E,
  Pfeiffer L N and West K W 2001 \emph{Phys. Rev.} B \textbf{63}
  121311(R)
  
\bibitem{Cronenwett} Cronenwett S M, Lynch H J, Goldhaber-Gordon D,
  Kouwenhoven L P, Marcus C M, Hirose K, Wingreen N S and Umansky V
  2002 \emph{Phys.\ Rev.\ Lett.} \textbf{88} 226805
  
\bibitem{Rokhinson} Rokhinson L P, Pfeiffer L N and West K W 2006
  \emph{Phys.\ Rev.\ Lett.} \textbf{96} 156602
  
\bibitem{Berggren1} Wang C-K and Berggren K-F 1996 \emph{Phys. Rev.} B
  \textbf{54} R14257
  
\bibitem{Berggren2} Wang C-K and Berggren K-F 1998 \emph{Phys. Rev.} B
  \textbf{57} 4552
  
\bibitem{Berggren3} Starikov A A, Yakimenko I I and Berggren K-F 2003
  \emph{Phys. Rev.} B \textbf{67} 235319
  
\bibitem{Spivak} Spivak B and Zhou F 2000 \emph{Phys. Rev.} B
  \textbf{61} 16730
  
\bibitem{Reilly} Reilly D J, Buehler T M, O'Brien J L, Hamilton A R,
  Dzurak A S, Clark R G, Kane B E, Pfeiffer L N and West K W 2002
  \emph{Phys.\ Rev.\ Lett.} \textbf{89} 246801
  
\bibitem{Lieb} Lieb E and Mattis D 1962 \emph{Phys. Rev.} \textbf{125}
  164
  
\bibitem{Wigner} Wigner E 1934 \emph{Phys. Rev.} \textbf{46} 1002
  
\bibitem{schulz} Schulz H J 1993 \emph{Phys.\ Rev.\ Lett.} \textbf{71}
  1864
  
\bibitem{fietereview} Fiete G A 2007 \emph{Rev. Mod. Phys.}
  \textbf{79} 801

\bibitem{Chaplik} Chaplik A V 1980 \emph{Pisma Zh. Eksp. Teor. Fiz.}
  \textbf{31} 275; 1980 \emph{JETP Lett.} \textbf{31} 252
 
\bibitem{Hasse} Hasse R W and Schiffer J P 1990 \emph{Ann.\ Phys.}
  \textbf{203} 419
  
\bibitem{Piacente} Piacente G, Schweigert I V, Betouras J J and
  Peeters F M 2004 \emph{Phys. Rev.} B \textbf{69} 045324
  
\bibitem{Ruzin} Glazman L I, Ruzin I M and Shklovskii B I 1992
  \emph{Phys.  Rev.} B \textbf{45} 8454
  
\bibitem{Hausler} H\"ausler W 1996 \emph{Z. Phys.} B \textbf{99} 551
  
\bibitem{Matveev} Matveev K A 2004 \emph{Phys.\ Rev.\ Lett.}
  \textbf{92} 106801
  
\bibitem{matveev-prb} Matveev K A 2004 \emph{Phys.\ Rev.} B
  \textbf{70} 245319
  
\bibitem{mfl-prl} Matveev K A, Furusaki A and Glazman L I 2007
  \emph{Phys.\ Rev.\ Lett.} \textbf{98} 096403
  
\bibitem{mfl-long} Matveev K A, Furusaki A and Glazman L I 2007
  \emph{Phys.\ Rev.} B \textbf{76} 155440
  
\bibitem{faddeev} Faddeev L D and Takhtajan L A 1981 \emph{Phys.
    Lett.}  \textbf{85A} 375

\bibitem{Haldane} Haldane F D M 1982 \emph{Phys. Rev.} B \textbf{25}
  R4925
  
\bibitem{Okamoto} Okamoto K and Nomura K 1992 \emph{Phys.\ Lett.}
  \textbf{169A} 433
  
\bibitem{eggert} Eggert S 1996 \emph{Phys. Rev.} B \textbf{54} R9612
  
\bibitem{majumdar-ghosh1} Majumdar C K and Ghosh D K 1969 \emph{J.\
    Math.\ Phys.} \textbf{10} 1388
  
\bibitem{majumdar-ghosh2} Majumdar C K and Ghosh D K 1969 \emph{J.\
    Math.\ Phys.} \textbf{10} 1399
  
\bibitem{Thouless} Thouless D J 1965 \emph{Proc.\ Phys.\ Soc.\ London}
  \textbf{86} 893
  
\bibitem{Roger} Roger M 1984 \emph{Phys. Rev.} B \textbf{30} 6432
  
\bibitem{Katano} Katano M and Hirashima D S 2000 \emph{Phys. Rev.} B
  \textbf{62} 2573
  
\bibitem{Voelker} Voelker K and Chakravarty S 2001 \emph{Phys. Rev.} B
  \textbf{64} 235125
  
\bibitem{ceperley} Bernu B, Candido L and Ceperley D M 2001
  \emph{Phys.\ Rev.\ Lett.} \textbf{86} 870
  
\bibitem{KRM} Klironomos A D, Ramazashvili R R and Matveev K A 2005
  \emph{Phys.\ Rev.} B \textbf{72} 195343
  
\bibitem{Fogler1} Fogler M M and Pivovarov E 2005 \emph{Phys.\ Rev.} B
  \textbf{72} 195344
  
\bibitem{Fogler2} Fogler M M and Pivovarov E 2006 \emph{J. Phys.:
    Condens.\ Matter} \textbf{18} L7
 
\bibitem{kmm} Klironomos A D, Meyer J S and Matveev K A 2006
  \emph{Europhys. Lett.} \textbf{74} 679
  
\bibitem{klironomos-unpub} Klironomos A D 2006 unpublished
 
\bibitem{kmhm} Klironomos A D, Meyer J S, Hikihara T and Matveev K A
  2007 \emph{Phys. Rev.} B \textbf{76} 075302
 
\bibitem{White} White S R and Affleck I 1996 \emph{Phys. Rev.} B
  \textbf{54} 9862
  
\bibitem{Hamada} Hamada T, Kane J, Nakagawa S and Natsume Y 1988
  \emph{J.\ Phys.\ Soc.\ Jpn.} \textbf{57} 1891
  
\bibitem{Tonegawa} Tonegawa T and Harada I 1989 \emph{J.\ Phys.\ Soc.\
    Jpn.}  \textbf{58} 2902
  
\bibitem{Chubukov} Chubukov A V 1991 \emph{Phys.\ Rev.}\ B \textbf{44}
  4693
  
\bibitem{Allen} Allen D, Essler, F H L and Nersesyan A A 2000
  \emph{Phys.  Rev.} B \textbf{61} 8871
  
\bibitem{Itoi} Itoi C and Qin S 2001 \emph{Phys. Rev.} B \textbf{63}
  224423
 
\bibitem{Danneau} Danneau R, Clarke W R, Klochan O, Micolich A P,
  Hamilton A R, Simmons M Y, Pepper M and Ritchie D A 2006
  \emph{Appl.\ Phys.\ Lett.} \textbf{88} 012107
  
\bibitem{klochan} Klochan O, Clarke W R, Danneau R, Micolich A P, Ho L
  H, Hamilton A R, Muraki K and Hirayama Y 2006 \emph{Appl.\ Phys.\
    Lett.}  \textbf{89} 92105
  
\bibitem{Danneau2} Danneau R, Klochan O, Clarke W R, Ho L H, Micolich
  A P, Simmons M Y, Hamilton A R, Pepper M and Ritchie D A 2008
  \emph{Phys.\ Rev.\ Lett.} \textbf{100} 016403

\bibitem{MML} Meyer J S, Matveev K A and Larkin A I (2007) \emph{Phys.
    Rev. Lett.} \textbf{98} 126404
  
\bibitem{MK-unpub} Meyer J S and Matveev K A 2007 unpublished

\bibitem{Tsvelik} Reyes S A and Tsvelik A M 2006 \emph{Phys. Rev.} B
  \textbf{73} 220405(R)
  
\bibitem{Mattis} Mattis D C 1965 \emph{Theory of Magnetism} (New York:
  Harper \& Row) Ch~9
  
\bibitem{Vaks-Larkin} Vaks G A and Larkin A I 1965 \emph{Zh.\ Eksp.\
    Teor.\ Fiz.} \textbf{49} 975; 1966 \emph{Sov.\ Phys.\ JETP}
  \textbf{22} 678
  
\bibitem{Polyakov} See, e.g., Polyakov A M 1987 \emph{Gauge Fields and
    Strings} (New York: Harwood Academic Publishers)
  
\bibitem{RG1} Muttalib K A and Emery V J 1986 \emph{Phys.\ Rev.\
    Lett.}  \textbf{57} 1370
  
\bibitem{RG2} Fabrizio M 1993 \emph{Phys. Rev.} B \textbf{48} 15838
  
\bibitem{Ledermann} Ledermann U and Le Hur K 2000 \emph{Phys. Rev.}  B
  \textbf{61} 2497
  
\bibitem{balents} Balents L 2000 \emph{Phys. Rev.} B \textbf{61} 4429
  
\bibitem{unpub} Sitte M, Rosch A, Meyer J S, Matveev K A and Garst M
  2008 in preparation
  
\bibitem{starykh} Starykh O A, Maslov D L, H\"ausler W and Glazman L I
  2000 \emph{Interactions and Quantum Transport Properties of Lower
    Dimensional Systems} (Lecture Notes in Physics vol~544) ed T
  Brandes (New York: Springer) p~37

\bibitem{LL-DOS-1} Dzyaloshinskii I E and Larkin A I 1973 \emph{Zh.
    Eksp. Teor. Fiz.}  \textbf{65} 411; 1974 \emph{Sov. Phys. JETP}
  \textbf{38} 202
  
\bibitem{LL-DOS-2} Luther A and Peschel I 1974 \emph{Phys. Rev.} B
  \textbf{9} 2911

\bibitem{exp1} Hew W K, Thomas K J, Farrer I, Anderson D, Ritchie D A
  and Pepper M 2008 \emph{Physica} E \textbf{40} 1645

\bibitem{exp2} Hew W K, Thomas K J, Pepper M, Farrer I, Anderson D,
  Jones G A C and Ritchie D A 2008 \emph{Phys. Rev. Lett.}
  \textbf{101} 036801

\bibitem{CNT} Deshpande V V and Bockrath M 2008 \emph{Nature Physics}
  \textbf{4} 314

\bibitem{qmc} Shulenburger L, Casula M, Senatore G and Martin R M 2008
  \emph{Phys. Rev.} B \textbf{78} 165303

\bibitem{baranger-0807.4292} Guclu A D, Umrigar C J, Jiang H and
  Baranger H U \emph{Preprint} arXiv:0807.4292

\bibitem{yacoby-double-wires1} Auslaender O M, Yacoby A, de Picciotto
  R, Baldwin K W, Pfeiffer L N and West K W 2002 \emph{Science}
  \textbf{295} 825

\bibitem{yacoby-double-wires2} Auslaender O M, Steinberg H, Yacoby A,
  Tserkovnyak Y, Halperin B I, Baldwin K W, Pfeiffer L N and West K W
  \emph{Science} \textbf{308} 88

\bibitem{Larkin-Shikin} Larkin I A and Shikin V B 1990 \emph{Phys.
    Lett.}  A \textbf{151} 335

\bibitem{Chklovskii} Chklovskii D B, Matveev K A and Shklovskii B I
  1993 \emph{Phys. Rev.} B \textbf{47} 12605
\end{thebibliography}
\end{document}